\documentclass[aps,epsfig,showkeys,letter,nofootinbib,superscriptaddress,eqsecnum,]{revtex4} %
\usepackage[dvips]{graphicx} 
\usepackage{amssymb}
\usepackage{bm}
\usepackage{rotate}
\usepackage{yfonts}

\newcommand{\be}{\begin{equation}}
\newcommand{\ee}{\end{equation}}
\newcommand{\bea}{\begin{eqnarray}}
\newcommand{\eea}{\end{eqnarray}}
       
\newcommand{\refeq}[1]{Eq.~(\ref{eq:#1})}          
\newcommand{\refeqs}[2]{Eqs.~(\ref{eq:#1})--(\ref{eq:#2})}          
          
\newcommand{\reffig}[1]{Fig.~\ref{fig:#1}}          
\newcommand{\reftab}[1]{Tab.~\ref{tab:#1}}

\newcommand{\refsec}[1]{Section~\ref{sec:#1}}          
\newcommand{\refapp}[1]{Appendix~\ref{app:#1}}

\renewcommand{\v}[1]{\mathbf{#1}}

\newcommand{\ph}{\varphi}

\newcommand{\vk}{\v{k}}

\newcommand{\eps}{\varepsilon}
\renewcommand{\d}{\delta}

\newcommand{\rhob}{\overline{\rho}_m}
\newcommand{\rhobn}{\overline{\rho}_{m,0}}

\newcommand{\Mpch}{\,{\rm Mpc}/h}
\newcommand{\iMpch}{\,h/{\rm Mpc}}
\newcommand{\Lbox}{L_{\rm box}}
\newcommand{\kNy}{k_{\rm Ny}}
\newcommand{\kmax}{k_{\rm max}}
\newcommand{\Msunh}{\,M_{\odot}/h}

\newcommand{\Om}{\Omega_m}

\newcommand{\OL}{\Omega_\Lambda}
\newcommand{\Orc}{\Omega_{\rm rc}}
\renewcommand{\L}{\Lambda}

\newcommand{\rhoDE}{\rho_{\rm DE}}

\newcommand{\ga}{\textswab{g}}
\newcommand{\halofit}{\texttt{halofit} }

\begin{document}

\title{Cosmological Simulations of Normal-Branch Braneworld Gravity}

\author{Fabian Schmidt}
\affiliation{Theoretical Astrophysics, California Institute of Technology M/C 350-17,
Pasadena, California  91125-0001, USA\footnote{Present address.}}
\affiliation{Department of Astronomy \& Astrophysics, The University of
Chicago, Chicago, IL 60637-1433}
\affiliation{Kavli Institute for Cosmological Physics, Chicago, IL 
60637-1433}

\begin{abstract}
We introduce a cosmological model based on the normal branch of DGP
braneworld gravity with a smooth dark energy component on the brane.
The expansion history in this model is identical to $\Lambda$CDM, thus
evading all geometric constraints on the DGP cross-over scale $r_c$.
This well-defined model can serve as a first approximation to more general braneworld
models whose cosmological solutions have not been obtained yet.
We study the formation of large scale structure in this model in the
linear and non-linear regime using N-body simulations for different
values of $r_c$. 
The simulations use the code presented in \cite{DGPMpaper} and solve
the full non-linear equation for the brane-bending mode in conjunction with
the usual gravitational dynamics. The brane-bending mode is attractive
rather than repulsive in the DGP normal branch, hence the sign of the modified gravity
effects is reversed compared to those presented in \cite{DGPMpaper}.

We compare the simulation results with those of ordinary $\Lambda$CDM simulations
run using the same code and initial conditions. We find that the matter
power spectrum in this model shows a characteristic enhancement peaking
at $k\sim0.7\iMpch$. We also find that the abundance of massive halos is 
significantly enhanced. Other results presented here include the density
profiles of dark matter halos, and signatures of the brane-bending mode
self-interactions (Vainshtein mechanism) in the simulations.
Independently of the expansion history,
these results can be used to place constraints on the DGP model and future
generalizations through their effects on the growth of cosmological structure.
\end{abstract}


\date{\today}

\maketitle

\section{Introduction}
\label{sec:intro}

Braneworld scenarios with infinite extra dimensions
have received much interest in recent years,
as a possible explanation for the observed accelerated expansion of
the universe \cite{Deffayet01,DeffayetEtal02b}, as well as a way to tackle the 
cosmological constant problem \cite{deGrav,deRham}. The simplest such model
is the Dvali-Gabadadze-Porrati (DGP) model \cite{DGP}. In this model, matter
lives on a four-dimensional brane embedded in five-dimensional Minkowski
space. The gravity action consists of a five-dimensional Einstein-Hilbert
term plus an ordinary, four-dimensional term localized on the brane.
Gravity thus is five-dimensional on large scales, and reduces to four-dimensional
General Relativity on small scales. The transition is given by the
\textit{cross-over scale} $r_c$, which is defined by the ratio of five- and
four-dimensional gravitational constants: $r_c = G^{(5)}/2G^{(4)}$.

There are two branches of homogeneous and isotropic solutions in this model,
corresponding to the two possible ways of embedding the brane in (asymptotic) five-dimensional
Minkowski space. The self-accelerated branch ({\it sDGP}) has received attention
since it leads to a late-time accelerated expansion of the universe without
any dark energy or cosmological constant, if $r_c$ is of order the present
Hubble horizon $c/H_0$. However, this branch of the model is plagued by a ghost
instability when perturbed around the de Sitter solution \cite{LutyEtal,NicRat,GregoryEtal}.
In addition, the expansion history predicted by the self-accelerating DGP
model \cite{Deffayet01} does not appear to fit observations (e.g., \cite{FangEtal}).
The other, so-called normal branch of the theory ({\it nDGP}), does not have
a ghost instability. However, it does not lead to self-acceleration, so that it
is necessary to add a cosmological constant (tension) on the brane
\cite{SahniShtanov,LueStarkman,GianantonioEtal,LombriserEtal}, or, more generally
a form of stress energy with negative pressure. 

Hence, we are interested in generalizing the normal-branch DGP model in a way that
allows it to pass expansion history constraints. 
Much work is ongoing to extend the DGP braneworld scenario 
\cite{deRham,galileon,Gabadadze09}, and the proposed models are generally expected to exhibit
an expansion history close to $\Lambda$CDM, while the phenomenology of the modification
to GR is expected to be similar to the normal branch DGP model. However,
full cosmological solutions have not been obtained yet. Braneworld-inspired
{\it parametrized} generalizations of DGP have been proposed 
in \cite{AfshordiEtal,KW}. On the one hand, this approach 
might well capture features of a yet-unknown full generalized braneworld model.
On the other hand, these parametrizations do not constitute a well-defined theory.

Here, we instead opt for considering a well-defined, albeit somewhat
contrived model based on DGP, which nevertheless exhibits some of
the features of a generalized braneworld model (see \refsec{braneworld} for
a discussion). Our model consists of normal-branch DGP gravity, with
a general, smooth quintessence-type dark energy component on the brane tuned
so that the resulting expansion history is identical to a $\Lambda$CDM model.
This model which we call \textit{nDGP+DE} thus satisfies all geometric constraints
from, e.g. the cosmic microwave background (CMB) and Supernovae, and leaves the
cross-over scale $r_c$ as an almost free parameter of the model.
Furthermore, the $r_c \rightarrow \infty$ limit of the theory is
precisely General Relativity with a cosmological constant.
Our main aim in this paper is to study the growth of structure in nDGP+DE
in the linear and non-linear regime.

The evolution of linear perturbations in DGP has been studied
in \cite{KoyamaMaartens,SongEtalDGP,CardosoEtal,ScI}. On scales smaller than the horizon
and the cross-over scale, the model can be described as an effective scalar-tensor
theory. The massless scalar degree of freedom corresponds to displacements of the brane
and is called the \textit{brane-bending mode}. In linear perturbation theory,
the effect of the brane-bending mode is simply to rescale the effective
gravitational constant for the dynamical potential (the time-time piece of the metric),
while the geodesics of photons (i.e. gravitational lensing) are not affected.
In the self-accelerating branch of DGP, the brane-bending mode is repulsive,
weakening the effective gravitational force, while it is attractive in the normal
branch.

As all viable modified
gravity models, DGP contains a non-linear mechanism to restore General Relativity
in high-density environments. This is achieved by a strong self-coupling of the
brane-bending mode, which becomes effective within a characteristic scale, the 
so-called Vainshtein radius \cite{Vainshtein72,DeffayetVainshtein02,KoyamaSilva}.
The Vainshtein radius depends on the mass considered as well as its configuration.
While the prefactor of the self-coupling is model-dependent,
the form of the coupling is generic and is expected to be universal to
braneworld models \cite{galileon,ScI}.
For typical structures in the
Universe, the Vainshtein radius is of cosmological scale. Hence it is necessary
to follow the brane-bending mode and its self-interactions when studying
the formation of structure in the Universe. In \cite{DGPMpaper}, we presented
simulations of the self-accelerating DGP model which self-consistently solve
for the brane-bending mode (see also \cite{ScII}). 
Here, we present simulations of two nDGP+DE models,
extending the studies of non-linear structure formation presented in \cite{DGPMpaper} 
to the normal branch of DGP, where the effects of modified gravity
have the opposite sign. 

N-body simulations of normal-branch braneworld
models have previously been presented in \cite{KW}. The key difference
to our simulations is that while we solve the full brane-bending mode
equation, \cite{KW} employed an approximation where
the modified forces are parametrized by an effective
density-dependent gravitational constant $G_{\rm eff}(\d\rho)$. Unfortunately,
as already pointed out in \cite{KW}, this approximation does not recover the 
correct large-distance behavior for the brane-bending mode, so that in
principle artefacts of the approximation might appear even on large scales.
Furthermore, the approximation will become worse the higher the resolution
of the simulation is. We present a quantitative comparison of our simulations 
with the $G_{\rm eff}(\d)$ approximation in \refapp{KW}.

In addition to the matter power spectrum and halo mass function, we show results on 
the density profiles of halos, and signatures of the Vainshtein mechanism in the
simulation results. 
The results presented in this paper together with \cite{DGPMpaper} can serve
as a basis for modeling non-linear structure formation in DGP.
A model of some of the results presented here and in \cite{DGPMpaper} 
in the context of spherical collapse and the halo model will be
the subject of a forthcoming paper. Such a model can then be used to constrain
DGP and more generalized braneworld models using the wealth of large scale
structure observations available (see \cite{fRcluster} for a first attempt
in the case of $f(R)$ gravity).

The paper is structured as follows. In \refsec{DGP}, we describe the nDGP+DE models in
detail, including the evolution of linear perturbations and the Vainshtein mechanism.
We describe the simulations in \refsec{sim}, and present the results
in \refsec{res}. We conclude in \refsec{concl}.

\section{DGP models}
\label{sec:DGP}

As a gravitational
framework for our nDGP+DE cosmology, 
we choose the normal branch of the DGP model \cite{Deffayet01}.
However,
we add a smooth, quintessence-type dark energy on the brane whose
equation of state is adjusted precisely to cancel the unwanted effects
of the extra dimension on the expansion history, yielding an
expansion history identical to $\Lambda$CDM. Hence, {\it independently}
of the expansion history, the cross-over scale $r_c$ is a free parameter
which is not constrained anymore to be of order $H_0^{-1}$ in this model.
The $r_c\rightarrow \infty$ limit of this theory is General Relativity
with a cosmological constant.

We will consider flat cosmologies throughout, and consider two cosmologies
of the nDGP+DE type: one with $r_c=500\:$Mpc (``nDGP--1''), and one
with $r_c=3000\:$Mpc (``nDGP--2''). We will also compare our results
with the cosmological simulations of the self-accelerating DGP
model without dark energy (``sDGP'') presented in \cite{DGPMpaper}.

\begin{figure}[t!]
\centering
\includegraphics[width=0.48\textwidth]{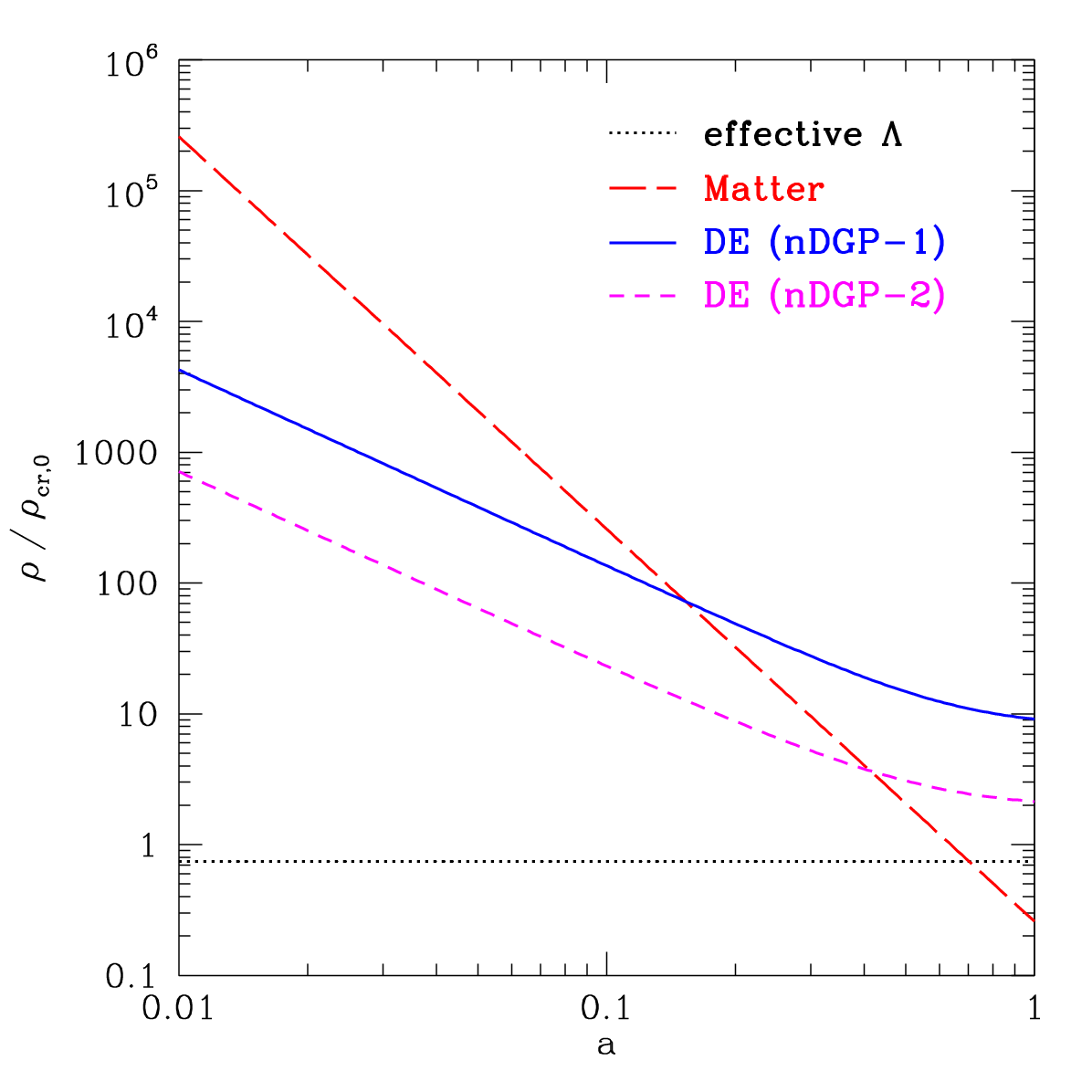}
\includegraphics[width=0.48\textwidth]{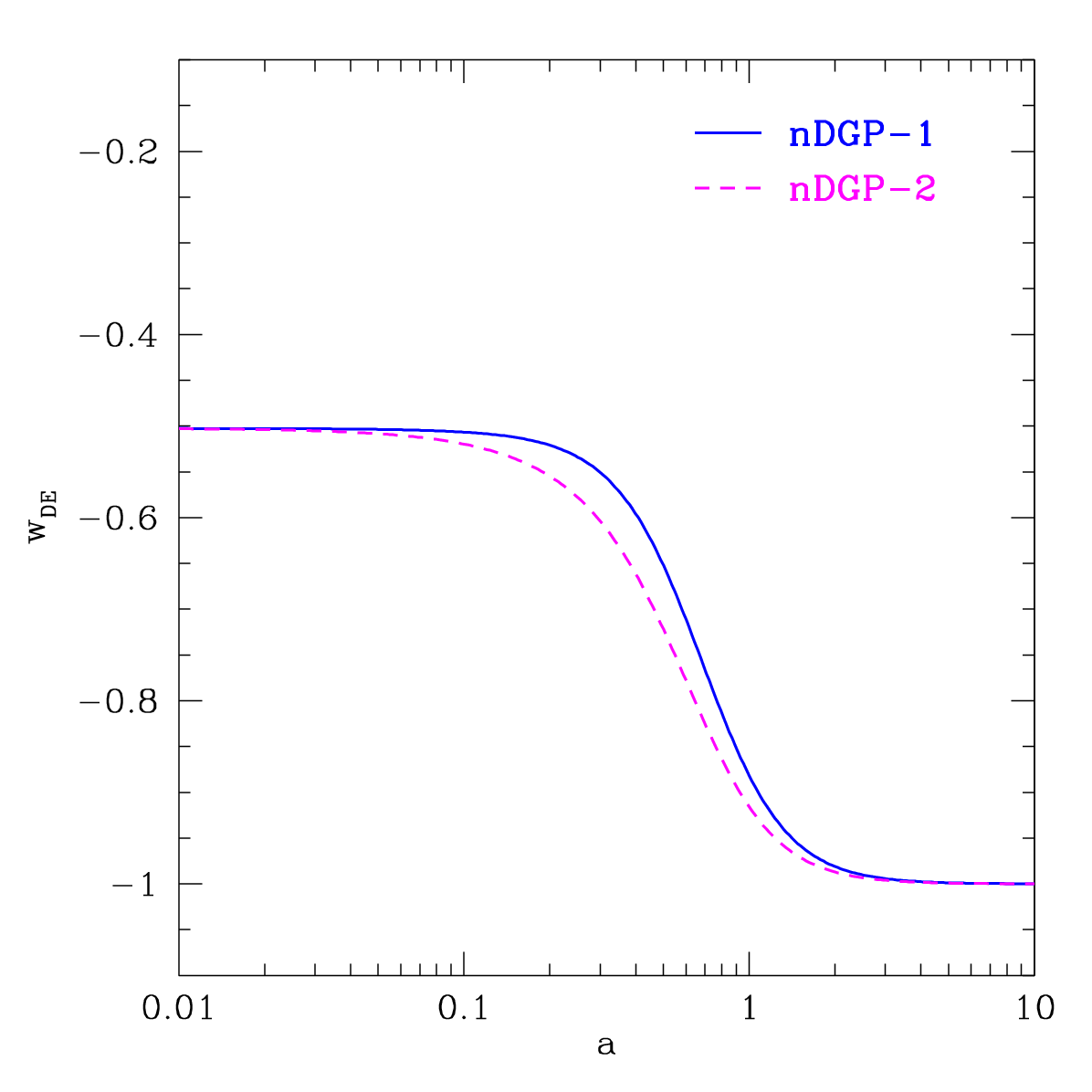}
\caption{{\small Behavior of the dark energy component in the
nDGP+DE models: density $\rhoDE$ and $\rho_m$, as well as the resulting
effective $\rho_\Lambda$ ({\it left panel});
equation of state $w_{\rm DE}$ of the dark energy ({\it right panel}).
The model parameters are defined in \reftab{params}.}
\label{fig:DE}}
\end{figure}
\subsection{Background evolution}
\label{sec:bg}

We start with the modified Friedmann equation for the normal branch of DGP
(e.g. \cite{SahniShtanov,LombriserEtal}):
\be
E(a)\equiv\frac{H(a)}{H_0} = \sqrt{\Om a^{-3} + \rho_{\rm DE}(a)/\rho_{\rm cr,0} + \Orc} - \sqrt{\Orc},
\label{eq:HDGP}
\ee
where $H_0$ is the Hubble rate $\dot a/a$ today, and
\be
\Orc \equiv \frac{1}{4H_0^2r_c^2};\quad \Omega_m \equiv \rhobn/\rho_{\rm cr,0};\quad \rho_{\rm cr,0} = \frac{3 H_0^2}{8\pi G}.
\ee
Here $\rhob$ and $\rho_{\rm DR}$ stand for the background energy density of 
matter and smooth
dark energy, respectively. We can now simply equate \refeq{HDGP} with the Friedmann
equation in GR for a flat $\L$CDM universe (with $\OL = 1-\Om$):
\be 
E_{\L\rm CDM}(a) = \sqrt{\Om a^{-3} + \OL}.
\label{eq:HLCDM}
\ee
and solve for the dark energy density. Note that in this model, one would infer
the {\it correct} value of $\Om$ from geometric and early-Universe tests,
such as BBN, CMB, SN, and $H_0$ measurements, which rely on the assumption
of the Friedmann equation \refeq{HLCDM}\footnote{Of course, the 
interpretation of $\L$ as a cosmological constant on the other hand would be very misguided.}.
This greatly simplifies the interpretation
of our results below.
We then obtain the following dark energy density:
\textbf{[Author's note: the following two equations are incorrect in the published version, as pointed out by \cite{bag/etal:2018}. I have corrected them here, and show the correct results in \reffig{DE}. None of the following results, nor those in followup papers, are affected since they simply use the expansion history given in \refeq{HLCDM}.]}
\be
\rhoDE(a) = \rho_{\rm cr,0} \left [ \OL + 2 \sqrt{\Orc} \sqrt{\OL + \Om a^{-3}}
  \right].
\ee
Note that this energy density is positive definite. In particular,
we can derive the following two limits:
\be
\rhoDE(a) = \rho_{\rm cr,0} \cdot \left \{ 
\begin{array}{ll}
  2\sqrt{\Orc \Om} a^{-3/2},
  & \Om a^{-3} \gg \Orc, \OL
  \vspace*{0.2cm}\\
  \OL + 2 \sqrt{\OL \Orc},
  & \Om a^{-3} \ll \Orc, \OL.
\end{array} \right .
\ee
This shows that the equation of state $w_{\rm DE}=p_{\rm DE}/\rhoDE$ is always
greater than -1: it approaches $-1/2$ for $a \ll 1$, and $-1$ for $a\gtrsim 1$
(\reffig{DE}). For $r_c \rightarrow \infty$, i.e. $\Orc \rightarrow 0$,
$\rhoDE = \OL\rho_{\rm cr,0} = $const., confirming the $\Lambda$CDM limit.
Note that in the matter-dominated regime, the dark energy 
density increases with decreasing cross-over scale $r_c$ (i.e., increasing 
$\Orc$), while at late times, when the DGP term dominates, it becomes
independent of $r_c$ and only depends on the effective $\L$CDM parameters.
\reffig{DE} shows the evolution of the
dark energy density and the equation of state for the two models we have
simulated, nDGP--1 ($r_c=500\:$Mpc) and nDGP--2 ($r_c=3000\:$Mpc).
Ref.~\cite{bag/etal:2018} provide a quintessential potential which very accurately realizes this time-dependent equation of state.

\begin{figure}[t!]
\centering
\includegraphics[width=0.48\textwidth]{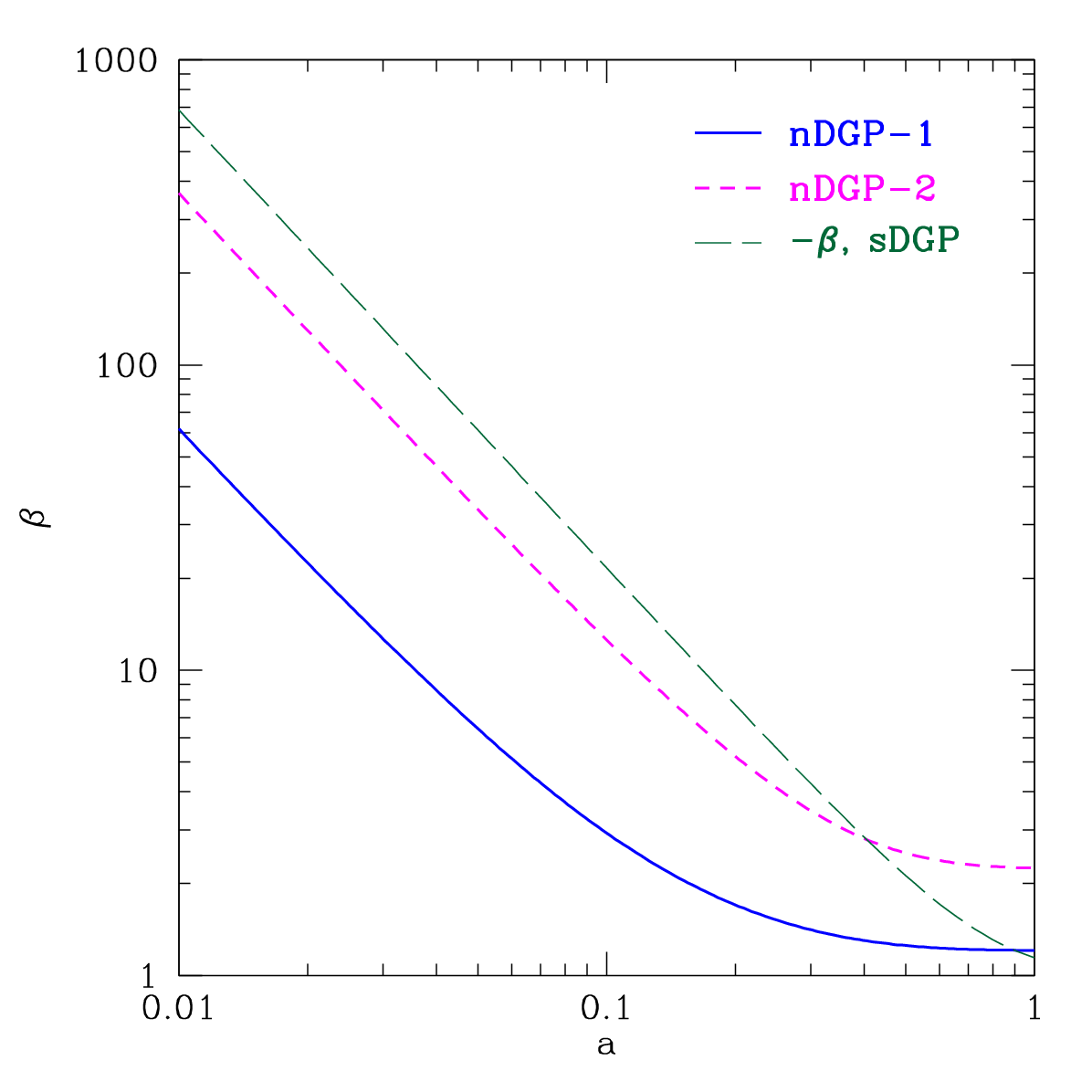}
\includegraphics[width=0.48\textwidth]{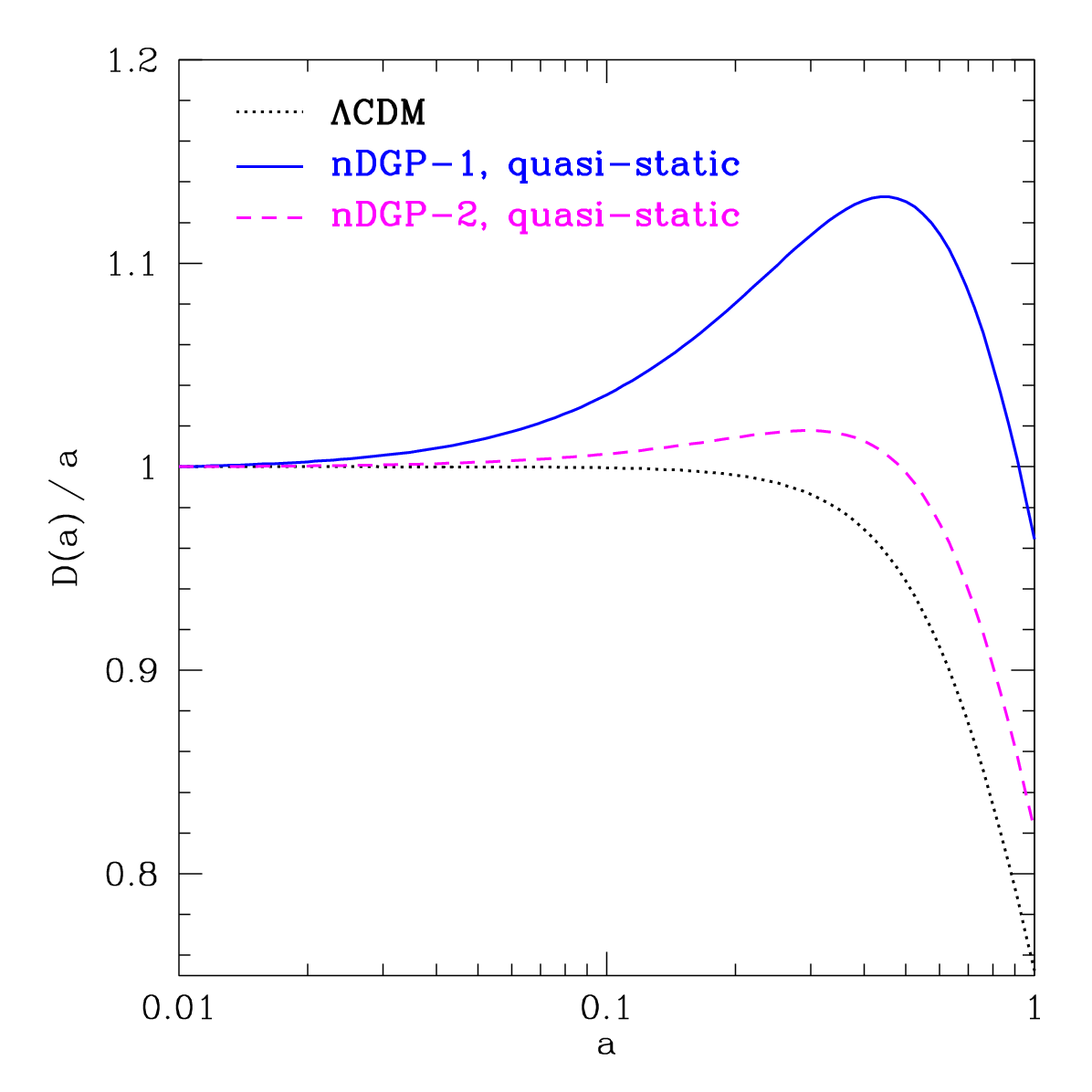}
\caption{{\small \textit{Left panel:} The $\beta$ function \refeq{beta} in the nDGP+DE models.
We also show the function for the self-accelerating DGP model considered
in \cite{DGPMpaper} (note that $\beta < 0$ for that model).
\textit{Right panel:} Linear growth factor $D(a)/a$ as a function of $a$, 
where $D(a)$ is normalized to $a$ in the matter-dominated
regime, for $\L$CDM and nDGP+DE models in the quasi-static regime
($k \gtrsim 0.01\iMpch$).}
\label{fig:betaD}}
\end{figure}

\subsection{Linear cosmological perturbations}
\label{sec:linear}

We now consider the evolution of cosmological perturbations in
our nDGP+DE model. We work in conformal Newtonian gauge and write
the metric as:
\be
ds^2 = -[1 + 2\Psi(\v{x},t)] dt^2 + [1 + 2\Phi(\v{x},t)] a^2(t) d\v{x}^2
\label{eq:metric}
\ee
Since we assume that the additional dark energy component is smooth,
the growth of structure in this model proceeds in the
same way as in the normal branch with brane tension \cite{LueStarkman,GianantonioEtal,LombriserEtal}, except that
the expansion history is altered. In particular, on small scales
$k \gtrsim 0.01\iMpch$, time derivatives may be neglected
with respect to spatial derivatives, which we call the quasi-static regime.
If, in addition, $k \gg r_c^{-1}$, DGP reduces to an effective scalar-tensor
theory, where the brane-bending mode $\ph$ plays the role of the
additional degree of freedom. The linearized brane-bending mode
equations then lead to a scale-independent but time-dependent 
modification of the
linear growth equation for matter perturbations, $\d = (\rho-\rhob)/\rhob$
(e.g. \cite{LueEtal04,SongEtalDGP,KoyamaSilva}):
\bea
\ddot \d + 2 H \dot \d &=& \frac{k^2}{a^2} \Psi,
\quad \left ( \; \frac{k}{a H} \gg 1 \;\right ) \nonumber\\
\frac{k^2}{a^2} \Psi &=& 4\pi G_{\rm lin}(a)\rhob \d, \label{eq:PsiQS}\\
G_{\rm lin}(a) &=& G\:\left(1 + \frac{1}{3\beta(a)} \right ),
\label{eq:growth}
\eea
where dots denote time derivatives, $G$ is Newton's constant, and,
in the normal branch of DGP:
\be
\beta(a) = 1 + 2 H(a)\, r_c \left ( 1 + \frac{\dot H(a)}{3 H^2(a)} \right ).
\label{eq:beta}
\ee
The function $\beta$ thus quantifies the departure from GR, which
is restored when $\beta \gg 1$ at early times (\reffig{betaD}, left panel).
In the normal branch, $\beta > 1$ and hence graviational forces are enhanced 
by up to 1/3. Conversely, in the self-accelerating branch $\beta < -1$
and forces are suppressed. Since we
only add an additional smooth stress-energy component and do not 
modify the gravity sector, the changes to standard DGP only come in through 
the expansion history in \refeq{beta}. 

\reffig{betaD} (right panel) shows the matter
growth factor $D(a) = \d(a)/\d(a_i)$ in the quasi-static regime as a function 
of $a$. The normalization is such that $D(a) = a$ in the matter-dominated
regime. As in GR+DE, the growth factor is independent of scale in the 
quasi-static regime ($k \gg r_c^{-1}$). In this case, the
growth is always enhanced in the nDGP+DE models with respect to $\L$CDM. 
On such scales much smaller than $r_c$, 
the five-dimensional nature of gravity is not important, and the effect
of the attractive brane-bending mode is the cause for the differences to $\L$CDM.

More generally, the effect of the brane-bending mode $\ph$ on the metric
potentials $\Psi$, $\Phi$ on sub-horizon and sub-$r_c$ scales is given by:
\bea
\Psi &=& \Psi_N + \frac{1}{2}\ph,\label{eq:Psi}\\
\Phi &=& -\Psi_N + \frac{1}{2}\ph.\label{eq:Phi}
\eea
Here, we have defined the ``Newtonian'' potential $\Psi_N$ via
the usual Poisson equation,
\be
\frac{k^2}{a^2} \Psi_N = 4\pi G\rhob \d. \label{eq:PsiN}\\
\ee
Note that the propagation of light is not directly affected in DGP, as is usually the
case in simple scalar-tensor theories: in the quasi-static regime, 
the lensing potential $\Phi_- \equiv (\Psi-\Phi)/2$ reduces to $\Psi_N$.
In other words, the relation between $\Phi_- $ determining the 
geodesics of photons and the matter overdensities is unchanged from GR \cite{LueEtal04}.

Finally, on larger scales approaching the horizon, time derivatives as well
as the effects of the extra dimension become important. Here, one can use the
parametrized post-Friedmann approach \cite{HuSawicki07}, calibrated
on calculations using the dynamical scaling solution \cite{SongEtalDGP},
or the full numerical solution \cite{CardosoEtal}.
This approach was also used in \cite{LombriserEtal} to place constraints
on DGP models with cosmological constant (brane tension), and we
adopt the parameters for the normal branch given in 
\cite{LombriserEtal}. Since we are considering models with $r_c$ smaller
than the present horizon scale, care has to be taken on scales significantly 
larger than $r_c$ where the PPF parametrization has not been calibrated
with the full 5D calculation. Such large scales are never relevant for
the simulations presented here, however.

Like all N-body simulations, our simulations work in the quasi-static
approximation. The full DGP equations contain non-local terms
involving $\sqrt{k^2}/r_c$ which could be relevant for small $r_c$.
We verify that neglecting these terms has a negligible impact on
the scales probed by the simulations in \refsec{sim}.

\subsection{Vainshtein mechanism}
\label{sec:vain}

If the modified force law given by $G_{\rm lin} \neq G$ was valid on 
all scales, the DGP model would predict order unity deviations from
the GR values of post-Newtonian parameters in the Solar System.
However, the brane-bending mode $\ph$ has self-interactions
which suppress its value in high-density regions. In the quasi-static
regime, the equation of motion can be written as:
\be
\nabla^2 \ph + \frac{r_c^2}{3\beta\,a^2} [ (\nabla^2\ph)^2
- (\nabla_i\nabla_j\ph)(\nabla^i\nabla^j\ph) ] = \frac{8\pi\,G\,a^2}{3\beta} \delta\rho,
\label{eq:phiQS}
\ee
where $\d\rho = \rhob\d$. The motion of particles is then governed by the dynamical potential
$\Psi$ as in \refeq{Psi}, which receives a contribution from $\ph$.
Linearizing 
\refeq{phiQS} together with \refeq{Psi} then yields \refeqs{PsiQS}{growth}.
For spherically symmetric masses, one can derive an analytic solution
to the field equation. Assuming a mass of constant density with radius $R$
(top hat), one obtains the following gravitational acceleration:
\bea
\ga &=& \ga_N + \frac{1}{2}\frac{d\ph}{dr} = \ga_N(r)\: [1 + \Delta(r)],
\label{eq:g_sph}\\
\Delta(r) &=&  \frac{2}{3\beta} \left \{ 
\begin{array}{rl}
r^3/r_*^3 \left ( \sqrt{1 + \left ( \frac{r_*}{r} \right )^3} - 1 \right ), & r \geq R \vspace*{0.2cm}\\
R^3/r_*^3 \left ( \sqrt{1 + \left ( \frac{r_*}{R} \right )^3} - 1 \right ), & r < R.
       \end{array} \right .
\label{eq:Delta}
\eea
where $\ga_N$ is the Newtonian acceleration of the spherical mass, and
 $r_*$ denotes a characteristic scale of the solution, the 
\textit{Vainshtein radius}, determined by $r_c$ and the Schwarzschild
radius of the mass $r_s$:
\be
r_*^3 = \frac{8\,r_c^2 r_s}{9\beta^2}.
\label{eq:rstar}
\ee
Close to the mass ($r \ll r_*$), force modifications are suppressed by $\sim (r/r_*)^{3/2}$.
At very large
distances, $r \gg r_*$, $\Delta(r)$ approaches the constant $1/(3\beta)$,
which exactly matches the linear solution, \refeq{growth}. 
Note that for the Sun and $r_c=3000$~Mpc, $r_* \sim 75$~pc, while
for the Milky way $r_*\sim$~Mpc. Hence, for $r_c\sim$~Gpc the non-linear 
interactions in \refeq{phiQS} are expected to be important on cosmological scales. Thus
it is crucial to self-consistently solve for the non-linearities in
$\ph$ in conjunction with the evolution of structure in the Universe.

Another limiting case in which \refeq{phiQS} is easily solvable is a plane
wave density field: in this case, the two non-linear terms cancel, and one
recovers the linearized solution. One might expect cosmological structure,
with its sheets and filaments as well as virialized structures, to lie
somewhere in between these limiting cases. Hence, the precise structure
of the non-linearity in \refeq{phiQS} is important for structure
formation.

\begin{figure}[t!]
\centering
\includegraphics[width=0.48\textwidth]{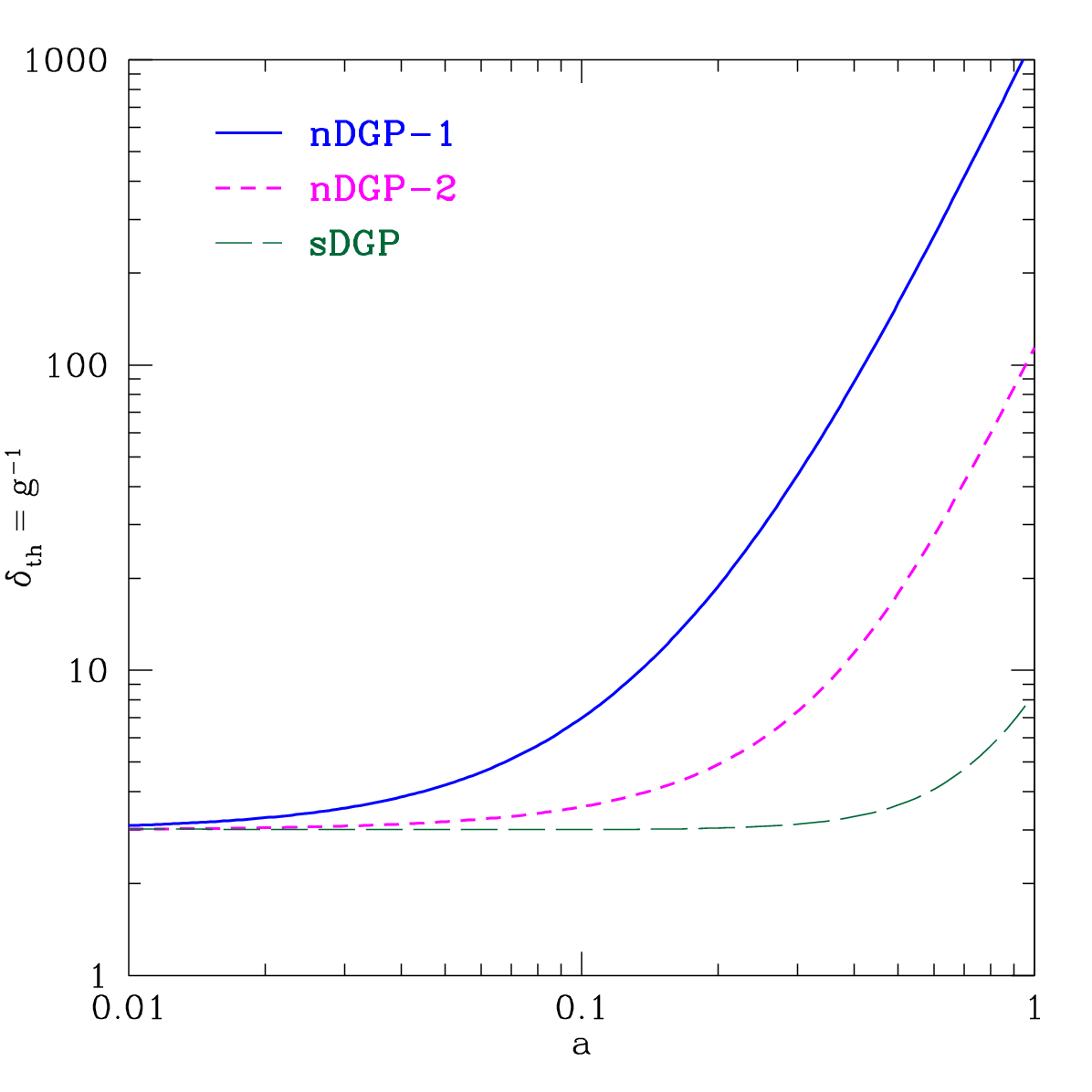}
\caption{{\small Overdensity threshold for the $\ph$ self-interactions,
defined as inverse of the non-linearity parameter $g$ [\refeq{NLest}],
as a function of scale factor in the different DGP cosmologies.}
\label{fig:gNL}}
\end{figure}

An order-of-magnitude estimate of where the coupling of the brane-bending mode becomes
strong can be obtained as follows. We take the ratio of the linear and non-linear
$\ph$ terms in \refeq{phiQS} and evaluate them for the linear solution $\ph_L$:
\be
\frac{r_c^2}{3\beta\,a^2}(\nabla^2\ph_L)^2 \left / \right . \nabla^2\ph_L 
= \frac{8\pi G\,r_c^2}{9\beta^2}\delta\rho
= \frac{H_0^2 r_c^2}{3\beta^2}\Omega_m a^{-3}\,\delta \equiv g\:\d,
\label{eq:NLest}
\ee
One can then estimate that wherever $\d \gtrsim \delta_{\rm th}\equiv g^{-1}$, 
the non-linear interactions
of $\ph$ are important (see \reffig{gNL}). The quantity $g$ defined here is 
the non-linearity parameter introduced in Sec. IV C of \cite{ScI}. At early 
times, $g\rightarrow 1/3$. By $z=0$, it decreases to $10^{-3}$ and $10^{-2}$ 
in the nDGP--1 and nDGP--2 cosmologies, respectively, and
$\sim 0.1$ in sDGP.
Today then, the non-linearity criterion will be satisfied for $\d \gtrsim O(1000)$ in the
case of nDGP--1, and $O(100)$ in the case of nDGP--2, which will only be the case within
the inner regions of dark matter halos. In the case of sDGP, $\delta_{\rm th}\approx 10$
today, which makes the non-linearities important in a significant fraction of the
Universe \cite{DGPMpaper}.

\subsection{Comparison with parametrized braneworld models}
\label{sec:braneworld}

The nDGP+DE model presented here is a full consistent (albeit somewhat contrived)
theory. It is worth comparing this model with the parametrized braneworld-inspired
models introduced in \cite{AfshordiEtal,KW}. On sub-horizon scales,
$k \gg a H,\,r_c$, these effective models yield the same equations as nDGP+DE
(\refsec{linear} and \ref{sec:vain}), except for a modified function $\beta$:
\be
\tilde\beta(a) = 1 + 2 (H r_c)^{2(1-\alpha)} \left ( 1 + \frac{\dot H}{3H^2} \right ).
\label{eq:betaKW}
\ee
For $\alpha=1/2$, this reduces to \refeq{beta}. Hence, the $\alpha=1/2$ model
of \cite{KW} is identical to our model on small scales.  Although the $\alpha=0$ model of \cite{KW}
has a somewhat different linear growth rate, the structure of the $\ph$ equation
is unchanged, and the behavior of the $\alpha=0$ model is correspondingly very
similar to that of the $\alpha=1/2$ case \cite{KW}.  Note that while the
simulated models of \cite{KW} are similar to ours, the simulation technique 
is different: the
simulations presented here solve the full $\ph$ equation~(\ref{eq:phiQS}),
while \cite{KW} adopted an approximation neglecting the tensorial structure
of \refeq{phiQS}. See \refapp{KW} for a discussion. 

On larger scales, $k \sim 1/r_c$, the full higher-dimensional theory becomes
important. Parametrized generalizations of DGP were proposed in \cite{AfshordiEtal},
which lead to interesting effects on the low multipoles of the CMB. In contrast,
the nDGP+DE model exhibits the five-dimensional features of the DGP model on
large scales, except of course for the expansion history which is $\Lambda$CDM.
Higher-dimensional effects however should not be important for predictions
on sub-horizon scales. In this regime, the results presented here for the nDGP+DE model will
be relevant to generalized braneworld models as well. 

\section{Simulations}
\label{sec:sim}

In \cite{DGPMpaper}, we have introduced cosmological
simulations of the self-accelerating DGP model, which self-consistently
solve the full non-linear \refeq{phiQS} of the brane-bending mode.
The code, based on an earlier version presented in \cite{HPMpaper}, 
uses a standard particle-mesh algorithm, augmented by a relaxation
solver for $\ph$. The Gauss-Seidel relaxation proceeds using Newton's method.
The convergence speed is greatly enhanced using multigrid techniques 
\cite{Brandt2,Briggs}.
We use this code for simulations of the nDGP+DE models, adapting the
expansion history and the $\beta$ [\refeq{beta}] function as it appears
in \refeq{phiQS}. See \cite{DGPMpaper} for details on the implementation
and tests of the code, which all apply in an analogous way to the
nDGP+DE simulations.

We simulate two nDGP+DE models which only differ in their value of $r_c$,
500~Mpc in the case of nDGP--1, and 3000~Mpc for nDGP--2.
The remaining
cosmological parameters defining the expansion history and the
primordial power spectrum are taken
from the best-fit flat $\L$CDM model given in \cite{FangEtal}. The data
used in their fit are the WMAP 5yr results, Supernova and $H_0$
measurements. \reftab{params} summarizes the cosmological parameters
of the simulations.

In addition to the full simulations, we performed simulations for
nDGP--1 and nDGP--2 using the linearized field equation for $\ph$, 
by replacing $G$ in the GR simulations with $G_{\rm lin}(a)$ given in 
\refeq{growth}. We refer to the latter
as \textit{linearized DGP} simulations. We also simulated
a standard $\L$CDM cosmology with identical expansion history.
Deviations of the nDGP+DE simulations from the $\L$CDM simulations
are thus purely due to the modification of gravity.
We have verified that our $\L$CDM simulations match fitting formulas
for the non-linear power spectrum and halo mass function.
Since these results are essentially the same as those shown in 
\cite{HPMpaperII,HPMhalopaper}, we do not show them again here.

\begin{figure}[t!]
\centering
\includegraphics[width=0.48\textwidth]{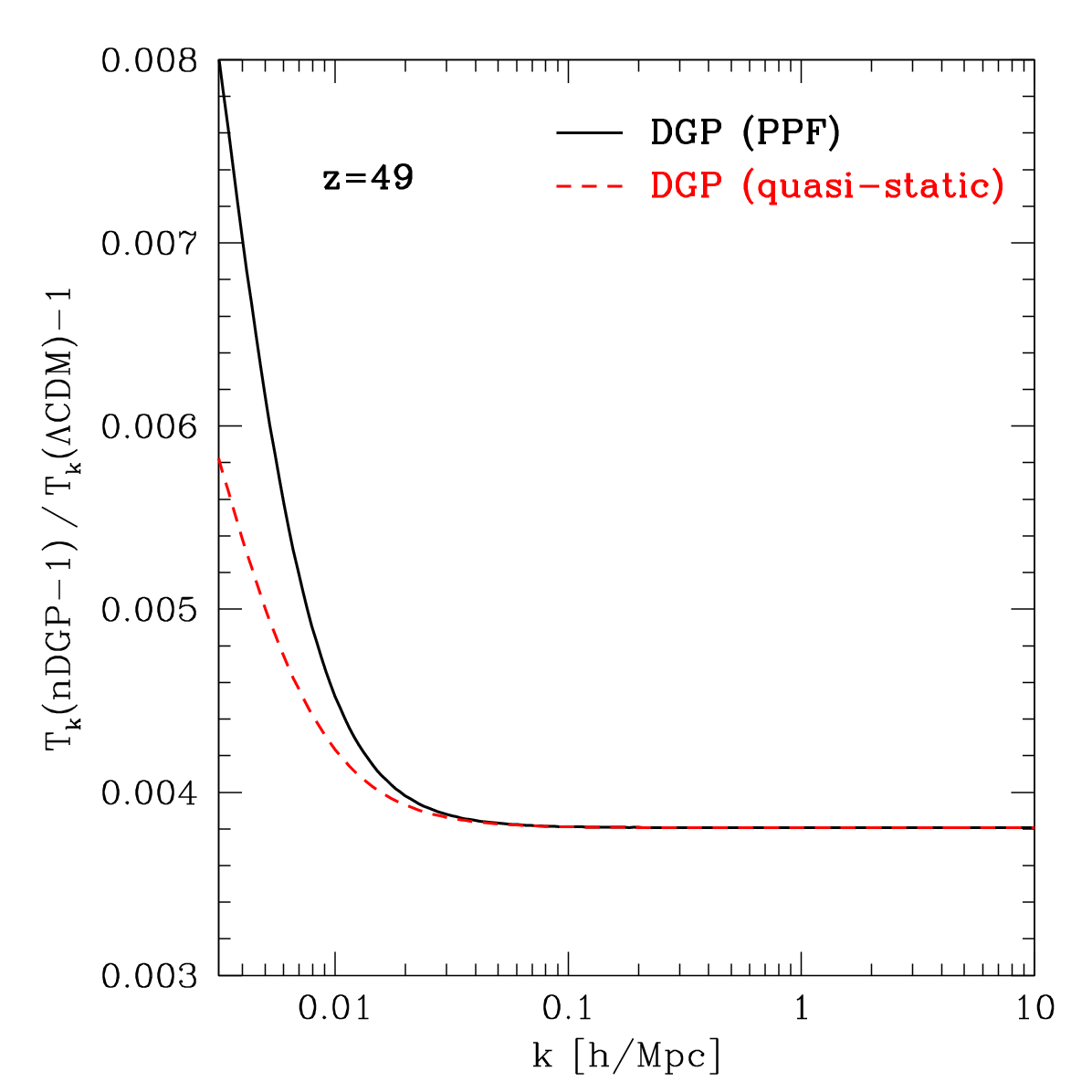}
\includegraphics[width=0.48\textwidth]{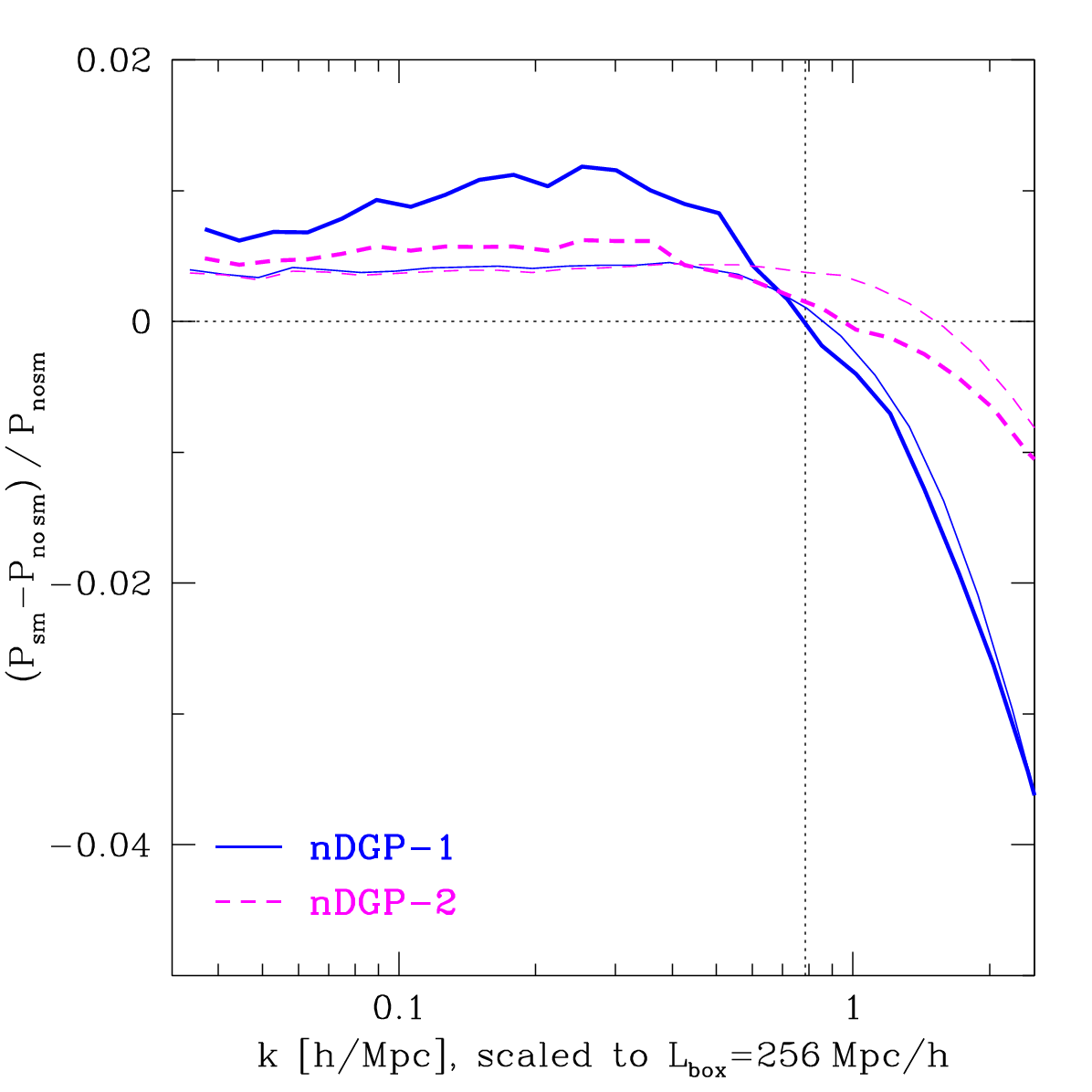}
\caption{{\small \textit{Left panel:} Deviation of the nDGP--1 matter transfer function 
from $\Lambda$CDM at the initial redshift of the simulations, $z_i=49$.
The solid line shows the PPF calculation for DGP which is used to correct 
the power spectrum for the initial conditions. The red dashed line
shows the same deviation calculated using the quasi-static calculation
[\refeqs{growth}{beta}].
\textit{Right panel:} Relative deviation of the matter power spectrum at $z=0$
measured in simulations with Gaussian smoothing
of the r.h.s of \refeq{phiQS} to simulations without smoothing.
The thick lines show results for $\Lbox=400\Mpch$, while the thin lines
show those for $\Lbox=128\Mpch$. All $k$ values were rescaled to those
for a $256\Mpch$ box. The vertical dashed line shows the maximum wavenumber
used in the analysis of the simulation results, $\kmax=\kNy/8$ (see text).}
\label{fig:ICsmooth}}
\end{figure}
\begin{table}[b!]
\begin{minipage}[t]{0.45\textwidth}
\caption{Cosmological parameters of the simulations.\label{tab:params}} 
\begin{center}
  \leavevmode
  \begin{tabular}{l|l|l|l}
\hline
 & $\L$CDM & nDGP--1 & nDGP--2 \\
\hline
$\Om = 1 - \OL$ & \  0.259 & \  0.259 & \  0.259 \\
$r_c$~[Mpc] & \  $\infty$ & \  500 & \  3000 \\
$\Orc$ & \  0 & \  17.5  & \  0.487 \\
$H_0$~[km/s/Mpc] & \  71.6 & \  71.6 & \  71.6 \\
\hline
$100\,\Omega_b\,h^2$ & \multicolumn{3}{|c}{2.26}\\
$\Omega_c\, h^2$ & \multicolumn{3}{|c}{0.110}\\
$\tau$ & \multicolumn{3}{|c}{0.0825}\\
$n_s$ & \multicolumn{3}{|c}{0.959}\\
$A_s$ ($k=0.05\,{\rm Mpc}^{-1}$) & \multicolumn{3}{|c}{$2.107\: 10^{-9}$}\\
\hline\hline
$\sigma_8(\Lambda\rm CDM)$\footnote{Linear power spectrum normalization today
of a $\Lambda$CDM model with the same primordial normalization.}
  & \multicolumn{3}{|c}{0.7892}\\
\hline
\end{tabular}
\end{center}
\end{minipage}
\hfill
\begin{minipage}[t]{0.45\textwidth}
\caption{Simulation type and number of runs per box size.\label{tab:runs}} 
\begin{center}
  \leavevmode
  \begin{tabular}{c|c c c c}
 & \multicolumn{3}{|c}{$L_{\rm box}$ [$\Mpch$]} \\ 
  \cline{2-5} 
 &\ \ $400$\ \ & $256 $\ \ \  & $128$\ \ \  & $64 $\ \ \  \\
\hline
   $\Lambda$CDM & 3 & 3 & 3 & 6\\
\hline
   Linearized nDGP--1 & 3 & 3 & 3 & 6\\
   Full nDGP--1 & 3 & 3 & 3 & 6\\
\hline
   Linearized nDGP--2 & 3 & 3 & 3 & 6\\
   Full nDGP--2 & 3 & 3 & 3 & 6\\
\hline\hline
$\kmax=\kNy/8$ [$h/\rm Mpc$] & 0.50 & 0.79 & 1.57 & 3.14 \\
\hline
$r_{\rm cell}$ [${\rm Mpc}/h$] & 0.78 & 0.50 & 0.25 & 0.13 \\
\hline
$M_{\rm min}$ [$10^{12}\Msunh$] & 219 & 57.3 & 7.17 & 0.90 \\
\hline
$r_s$ [grid cells]\footnote{Gaussian smoothing radius for  
full DGP simulations.} & 0.8 & 0.8 & 0.8 & -- \\
\hline
\end{tabular}
\end{center}
\end{minipage}
\end{table}

The simulations were started at $z_i=49$.  In order to test whether
this is early enough to capture the non-linear evolution in our
smallest box, we compared the power spectrum in that box after 10 time
steps ($a=0.04$) to the initial power spectrum ($a=0.02$).  The departures
from scale-free linear evolution in the power spectrum are below 1\% even
up to $k=2\kmax \approx 6\iMpch$. We conclude that given the limited
resolution of our simulations, there is no need to start the simulations
at an earlier time.

However, as can be seen from
\refeq{growth}, there are small residual effects from DGP even at
the initial epoch, where $\beta(z_i)$ is of order a few hundred.
In order to take into account these small modified gravity effects,
we correct the linear $\L$CDM 
power spectrum at $z_i$ as given by CAMB \cite{CAMB} by a factor $f^2(k)$
where $f(k) = T_{\rm PPF,DGP}(k,z_i)/T_{\Lambda\rm CDM}(k,z_i)$. Here,
$T_{\rm PPF,DGP}$ is the DGP transfer function calculated in the parametrized 
Post-Friedmann approximation \cite{HuSawicki07}, which has been shown
to be accurate at the few percent level on large scales. On small,
quasi-static scales the calculation is exact. $T_{\Lambda\rm CDM}$ was
calculated in the same way without any modification to gravity.
\reffig{ICsmooth} (left panel, solid line) shows the correction factor $f$
obtained in this way as a function of $k$. At $z_i=49$, the DGP effects on 
the transfer function are at the level of few $10^{-3}$. Hence, a
percent-level uncertainty on the correction will have a negligible impact
on the initial power spectrum.
Furthermore, for simplicity and because it is a small
correction, we run all simulations including $\L$CDM with the same
initial conditions corrected for nDGP--1 ($r_c=500$~Mpc).

\reffig{ICsmooth} (left panel) also shows the same correction
calculated in the pure quasi-static approach [\refeq{growth}], which
neglects the non-local $\sqrt{k^2}/r_c$ term in the Poisson equation
\cite{KW,ScI}.
On scales relevant for the simulations, the quasi-static approximation
is accurate. This is important, since the N-body simulation assumes a 
quasi-static regime. \reffig{ICsmooth} shows that this is justified 
at $z_i=49$ for the nDGP models considered here (the deviations from
the quasi-static assumption
will continue to decline at later times).

Our simulations use a 512$^3$ grid with 512$^3$ particles. The 
high number of particles was chosen in order to reduce the shot
noise in the density field. Since \refeq{phiQS} is non-linear in the
highest derivatives, it responds sensitively to small-scale inhomogeneities
in the density field. Increasing the number of particles helps in 
reducing the residual errors in the numerical solution of
\refeq{phiQS} to an acceptable level \cite{DGPMpaper}. Note that we
adopt a very strict (conservative) convergence criterion, in demanding
that the dimensionless RMS residuals of the field equation are less 
than $10^{-10}$.
For the self-accelerating simulations presented in \cite{DGPMpaper},
we employed a Gaussian smoothing of the r.h.s. of
\refeq{phiQS}. Fortunately, the convergence properties of the nDGP
models considered here are better, since the cross-over scale is smaller:
this raises the density threshold for which the non-linearities in $\ph$
become important. Correspondingly, particle noise in the density field
has less impact. This can also be seen by considering the Vainshtein
radius of a single particle in the simulations: for the self-accelerating
model of \cite{DGPMpaper}, $r_*$ was of order 1 grid cell. For the models
considered here, it is less than 0.5 grid cells. 

For this reason, we were able to eliminate the smoothing of the r.h.s. 
for our smallest box
($64\Mpch$ comoving size), where the smoothing effects are largest
\cite{DGPMpaper}. In addition, we lowered the smoothing radius
to 0.8 grid cells for all other boxes. These choices were again
made based on achieving an RMS residual less than $10^{-10}$ at all time
steps. We quantified
the effect of the smoothing by comparing full DGP simulations
with and without smoothing, where the latter slightly violate our 
upper bound on the residuals (maximum residual of $\sim 3-6 \times 10^{-10}$). 
\reffig{ICsmooth} (right panel) shows the power spectrum
at $z=0$ of simulations with smoothing relative to simulations without
smoothing using the same initial conditions. The effects of the smoothing
mainly appear on wavenumbers larger than the adopted maximum wavenumber,
given by $\kmax=\kNy/8$, where $\kNy$ is the Nyquist frequency of the
grid \cite{DGPMpaper}. Below $\kmax$, the differences are below
1.5\%, significantly smaller than our statistical uncertainties
on the final power spectrum measurement (\refsec{Pk}). The same was found
for the halo mass function.
Hence, no correction for smoothing effects is necessary.

Note that since the smoothing damps power
in the brane-bending mode on small scales, and the brane-bending mode
is attractive in the nDGP model, the simulations with smoothing of the
r.h.s. have suppressed power on small scales. \reffig{ICsmooth} also
shows that the power spectrum is stable even for simulations that
violate our adopted convergence criterion by a factor of 3-6.
We simulated four different comoving box sizes, from $400\Mpch$ to
$64\Mpch$. The number of runs for each model and box is summarized
in \reftab{runs}.

\section{Results}
\label{sec:res}

\subsection{Matter Power Spectrum}
\label{sec:Pk}

\begin{figure}[t!]
\centering
\includegraphics[width=0.48\textwidth]{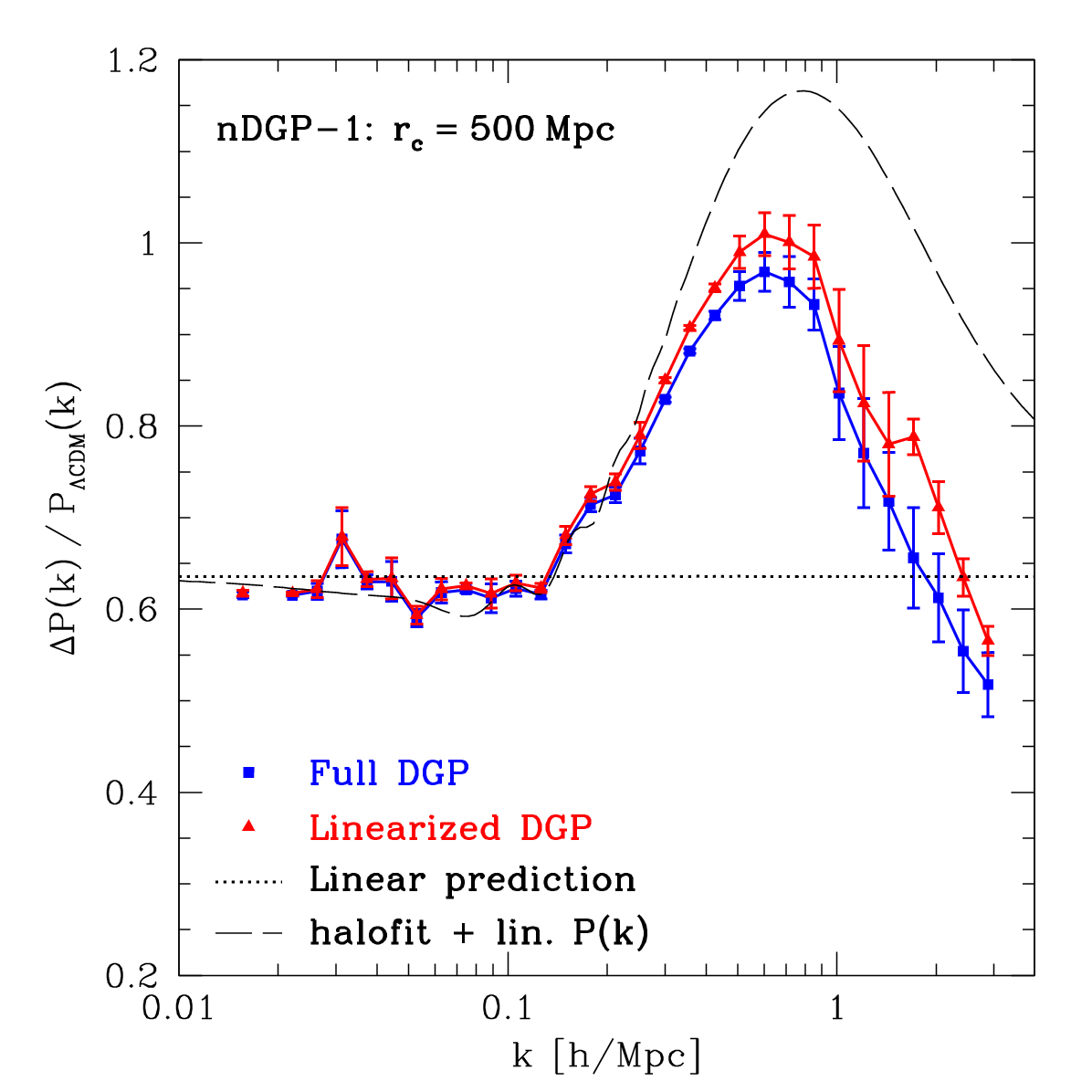}
\includegraphics[width=0.48\textwidth]{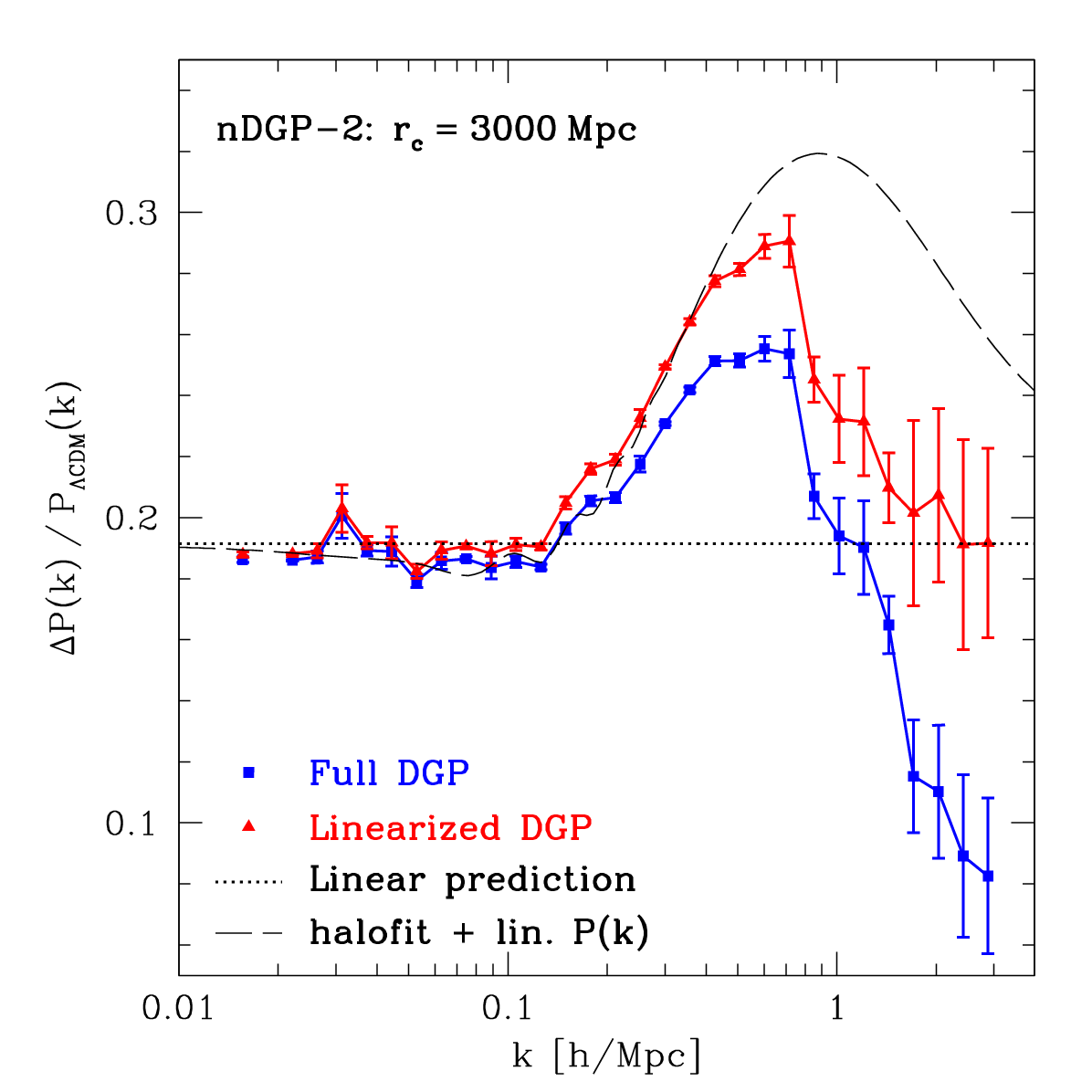}
\caption{{\small Relative deviation in the matter power spectrum from $\L$CDM at
$z=0$, $\Delta P(k)/P_{\Lambda\rm CDM}(k)\equiv (P_{\rm DGP}(k)-P_{\Lambda\rm CDM}(k))/P_{\Lambda\rm CDM}(k)$ for nDGP--1 \textit{(left panel)} and nDGP--2
\textit{(right panel)}. The blue squares and red triangles show the measurements
from the full and linearized DGP simulations, respectively. The short-dashed line 
shows the almost scale-invariant
deviation predicted using linear perturbation theory (\refsec{linear}),
while the long-dashed line shows the deviation obtained when using the
\halofit prescription together with the linear prediction for DGP.}
\label{fig:Pk}}
\end{figure}

First, we compare the power spectrum measured in DGP simulations $P_{\rm DGP}(k)$ to that
of the GR+DE simulations with the same expansion history, $P_{\Lambda\rm CDM}(k)$ (\reffig{Pk}). The
procedure is the same as that employed in \cite{HPMpaperII,DGPMpaper}: we
calculate the relative deviation of the power spectrum run by run, comparing
simulations with identical initial conditions. We then determine the
average deviation and its error by bootstrapping over realizations. In this way we are
able to reduce the variance of the deviations considerably.

The left panel in \reffig{Pk} shows the results for full and linearized simulations
of the nDGP--1 cosmology at redshift 0. Both simulations approach the scale-invariant
linear predictions
on the largest scales ($k \lesssim 0.1\iMpch$). For nDGP--1 on non-linear 
scales, we find good overall agreement with the results of Khoury \& Wyman
(Fig. 11 in \cite{KW}). This is expected, since we found that the effects on 
the power spectrum of the $G_{\rm eff}$--approximation used in \cite{KW} are at the level
of 5--10 percent in our moderate resolution simulations (\refapp{KW}).
The overall scale-dependence of the deviations in the non-linear matter power spectrum shows the typical behavior when comparing a cosmology with
the same linear power spectrum shape but slightly different normalization. 
The deviations
initially grow towards smaller scales, peak around $k\sim 0.7\iMpch$, and then decline 
again towards even smaller scales. At the peak, the matter power spectrum in nDGP--1
is enhanced by a factor of 2 with respect to $\Lambda$CDM, while the enhancement is
about 65\% in the linear regime.
In the halo model, the largest deviations occur at a scale 
roughly corresponding to the transition region between two-halo and one-halo terms.
On smaller scales, the power spectrum mainly probes the density profiles of dark
matter halos. If the halo profiles are not strongly affected by the increased growth, the 
deviations are expected to decrease again in this regime (see \refsec{prof} for a 
study of halo profiles). These trends are captured qualitatively by the \halofit
prescription \cite{halofit}, which is used to map  a linear power 
spectrum into a non-linear one and is calibrated on GR N-body simulations. 
The dashed line in \reffig{Pk} shows the deviation
of the non-linear $P(k)$ for DGP from that for $\Lambda$CDM obtained using \halofit
with the corresponding linear power spectra.
While the \halofit predictions reproduce the qualitative behavior, the overall magnitude of the 
deviations on non-linear scales is not matched. Interestingly, \halofit does
predict the slight suppression of the deviations {\it below} the linear prediction
on quasi-linear scales $k \lesssim 0.1\iMpch$. A modification of the \halofit parameters
to improve the fit to modified gravity simulations was presented in \cite{KW}.
Note that if there was a unique
prescription to go from linear power spectrum to the non-linear one \textit{at a 
fixed redshift}, then \halofit should be able to describe the linearized DGP
simulations, as the linear power spectrum in the nDGP+DE model is equivalent to 
a $\L$CDM power spectrum with a slightly higher normalization on these scales
($k \gtrsim 0.01\iMpch$). The discrepancy between the linearized DGP simulations and
the \halofit prediction shows that the linear-to-nonlinear mapping does depend
on the growth {\it history} (which is different in nDGP+DE), rather than just the 
growth factor at the present time. 

For nDGP--1, the deviations from GR
in the full simulations are only marginally suppressed compared to those in the
linearized simulations, on the scales accessible to our simulations. This is somewhat
expected, since the threshold overdensity for the Vainshtein mechanism in nDGP--1
is $\sim 1000$ (\refsec{vain}), which is only reached in the cores
of the most massive halos in our simulations.
The situation is slighly different in nDGP--2 (right panel of \reffig{Pk}): here
the effects of the non-linear interactions of $\ph$ are noticeable for
$k \gtrsim 0.2\iMpch$, and become increasingly important towards smaller scales.
The behavior as function of $k$ of the deviation in non-linear matter power is very similar
to that of nDGP--1, with the overall magnitude being smaller ($20-30$\%).

Note that in both nDGP cosmologies, the power spectrum deviation becomes significantly scale-dependent
at around $k\sim 0.1\iMpch$, in the region where the second and higher harmonics of the
baryon acoustic oscillations (BAO) are located. This scale-dependence should be taken into
account when using (sufficiently precise) BAO measurements to constrain DGP and other braneworld 
models.

\subsection{Halo Mass Function}
\label{sec:dndm}

\begin{figure}[t!]
\centering
\includegraphics[width=0.48\textwidth]{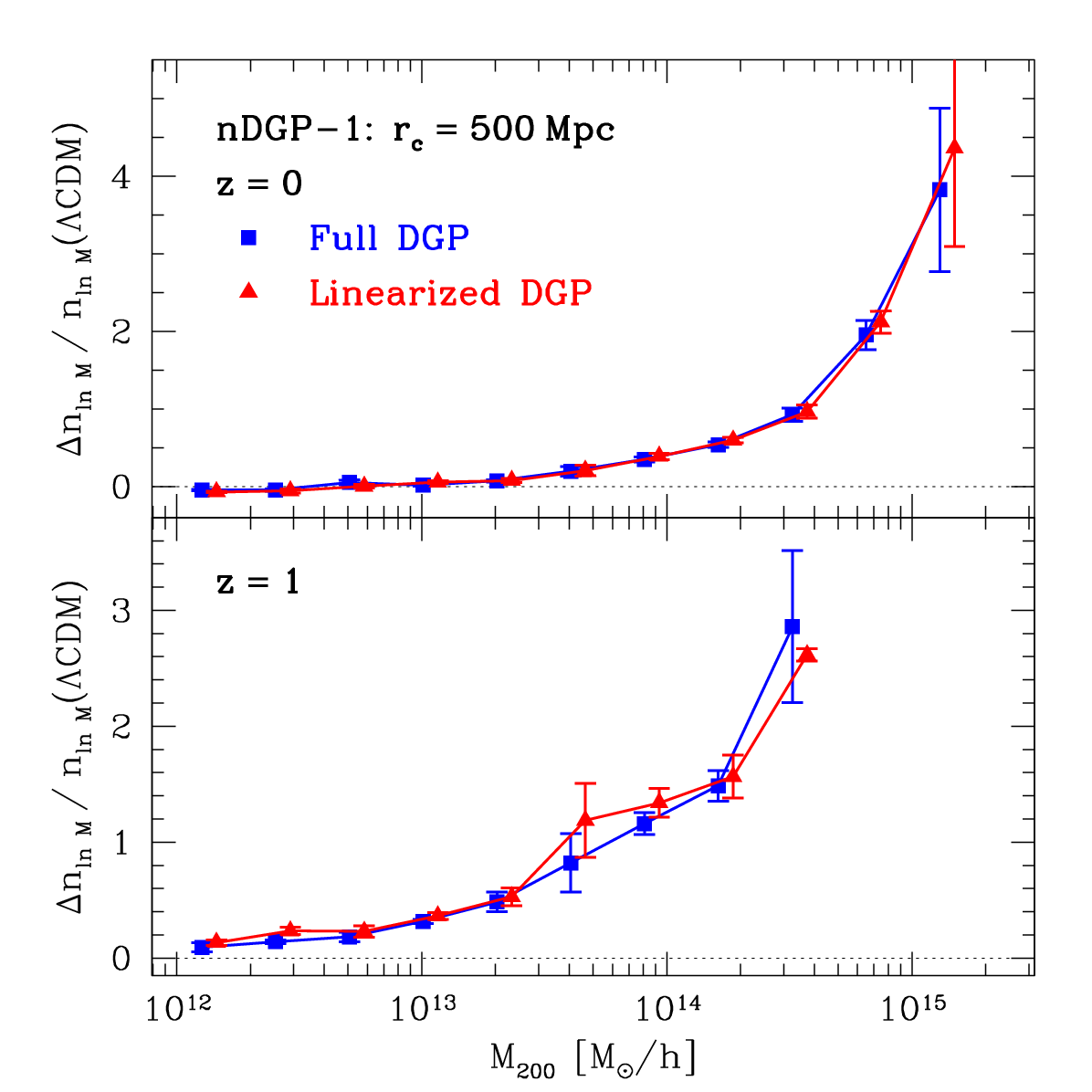}
\includegraphics[width=0.48\textwidth]{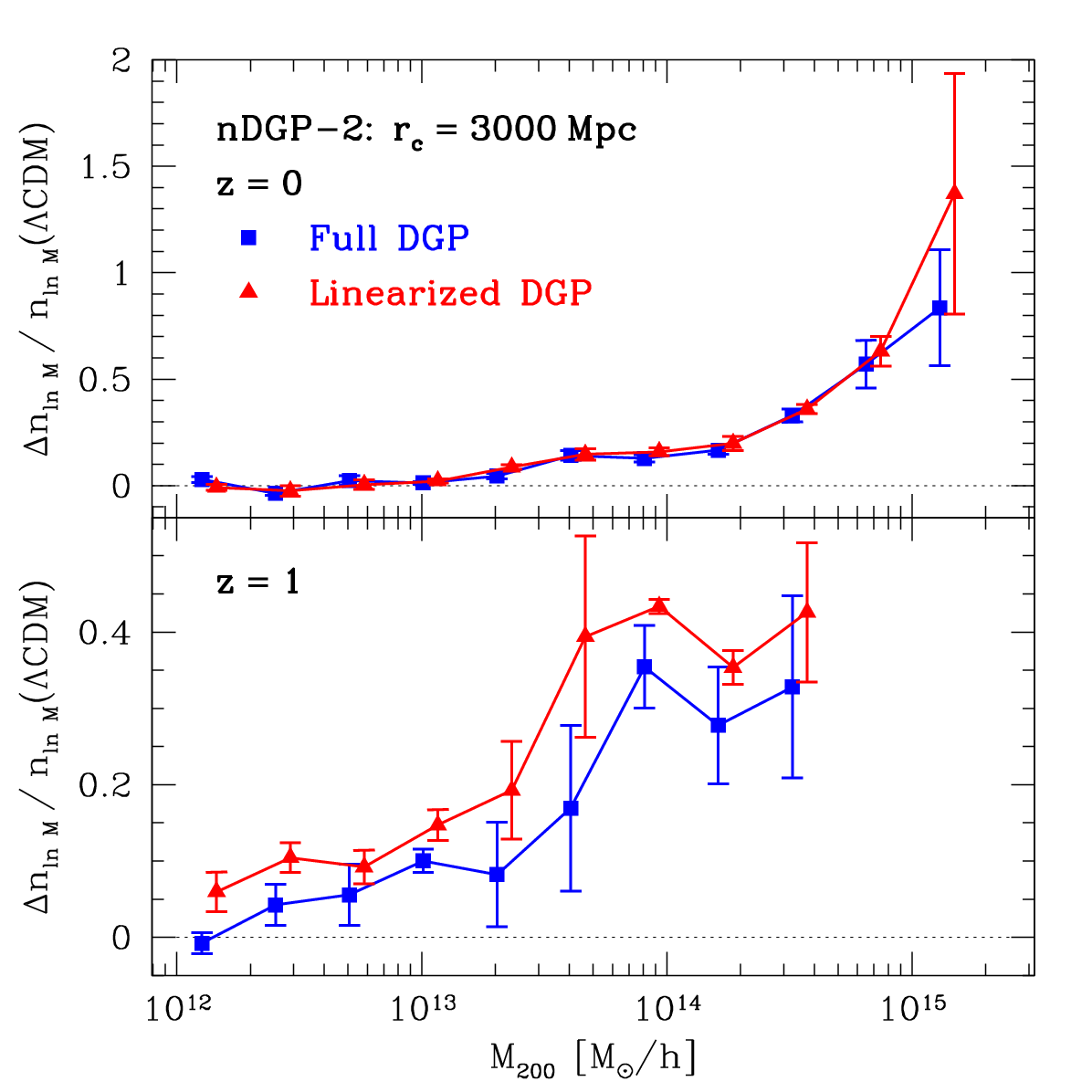}
\caption{{\small Relative deviation of the halo mass function from $\L$CDM,
$\Delta n_{\ln M}/n_{\ln M}(\Lambda\rm CDM)$ where $\Delta n_{\ln M}\equiv n_{\ln M}({\rm DGP})-n_{\ln M}(\Lambda\rm CDM)$, at
$z=0$ ({\it top}) and $z=1$ ({\it bottom}) for nDGP--1 \textit{(left panel)} 
and nDGP--2 \textit{(right panel)}. Blue squares and red triangles denote
the full and linearized DGP simulations, respectively.}
\label{fig:dndm}}
\end{figure}

Previous studies have shown that the abundance of dark matter halos is a sensitive
probe of modifications of gravity \cite{HPMhalopaper,DGPMpaper,ScII}. We identify
dark matter halos in the DGP and GR+DE simulations using a spherical overdensity
halo finder as described in \cite{HPMhalopaper}, with the same mass thresholds
as adopted in \cite{DGPMpaper}, corresponding to a minimum of 6400 particles
in a halo (see \reftab{runs} for the minimum halo mass for each
box size). The halo mass $M_{200}$ is defined
as the mass contained within a spherical region of radius $r_{200}$ around the halo 
center of mass,
whose average density is $200\times\rhob$. We calculate the mass function deviation
of the nDGP simulations from the GR simulations run-by-run, and then take the
average, as done for the power spectrum.

\reffig{dndm} shows the relative deviation of the mass function
$n_{\ln M} \equiv dn/d\ln M$ in DGP from $\Lambda$CDM, where $n(M)$ is 
the number density of halos of a given mass. The left panel shows the result
for nDGP--1, the right panel for nDGP--2.
In each case, the top panel shows the deviation at $z=0$, while
the bottom panel shows $z=1$. The qualitative behavior is the same in all cases
and is expected: the increase in halo abundance grows rapidly with mass, as
the abundance of massive halos is exponentially sensitive to the amplitude
of fluctuations today, which is increased in the nDGP cosmologies. 
Since the abundance of massive halos can be probed using galaxy clusters, this observable
can serve as a sensitive probe of braneworld gravity.
Current observations should be able to put a lower limit on the cross-over scale $r_c$
in the Gpc range, a constraint which would be independent of the specific
DGP expansion history.

Note that
at a fixed mass, the relative enhancement of the mass function at $z=1$ is comparable
to the one at $z=0$. This is due to a cancelation of two effects: the growth enhancement
in nDGP is smaller at $z=1$ than at $z=0$; on the other hand, a fixed mass
corresponds to a rarer fluctuation at $z=1$ than at $z=0$, hence increasing the sensitivity
to the growth enhancement. 

At $z=0$, the Vainshtein mechanism does not strongly affect the halo mass 
functions in our 
nDGP+DE simulations. For nDGP--2, there is an indication of a slightly suppressed abundance
of the highest mass halos in the full simulations compared to the linearized simulations,
which one expects since the most massive halos have the largest effective Vainshtein radii.
This situation changes somewhat at higher redshifts: in nDGP--2 at $z=1$, the mass function in the 
full simulations is lower by $5-10$\% at all masses. This can be explained by the lower
threshold overdensity at which the $\ph$ self-coupling becomes important (\refsec{vain}):
in nDGP--2, the non-linearity parameter $g$ is  $\sim 0.06$ at $z=1$, compared to $0.01$ at $z=0$.
Since a much larger portion of the universe has $\d \gtrsim 10$ rather than $\d \gtrsim 100$, the
self-coupling of $\ph$ affects a broad range of halo masses at $z=1$.


\subsection{Halo Profiles}
\label{sec:prof}

\begin{figure}[t!]
\centering
\includegraphics[width=0.32\textwidth]{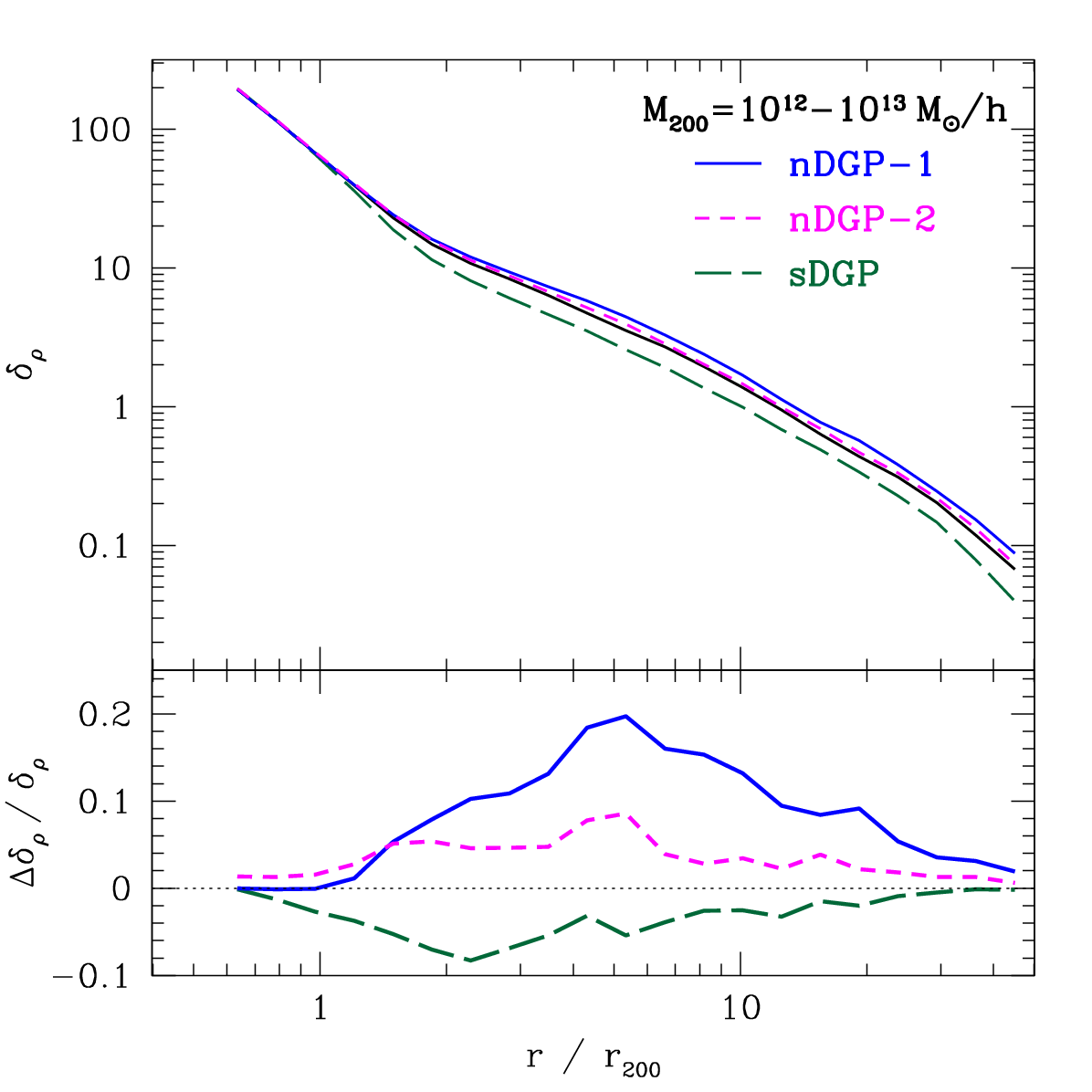}
\includegraphics[width=0.32\textwidth]{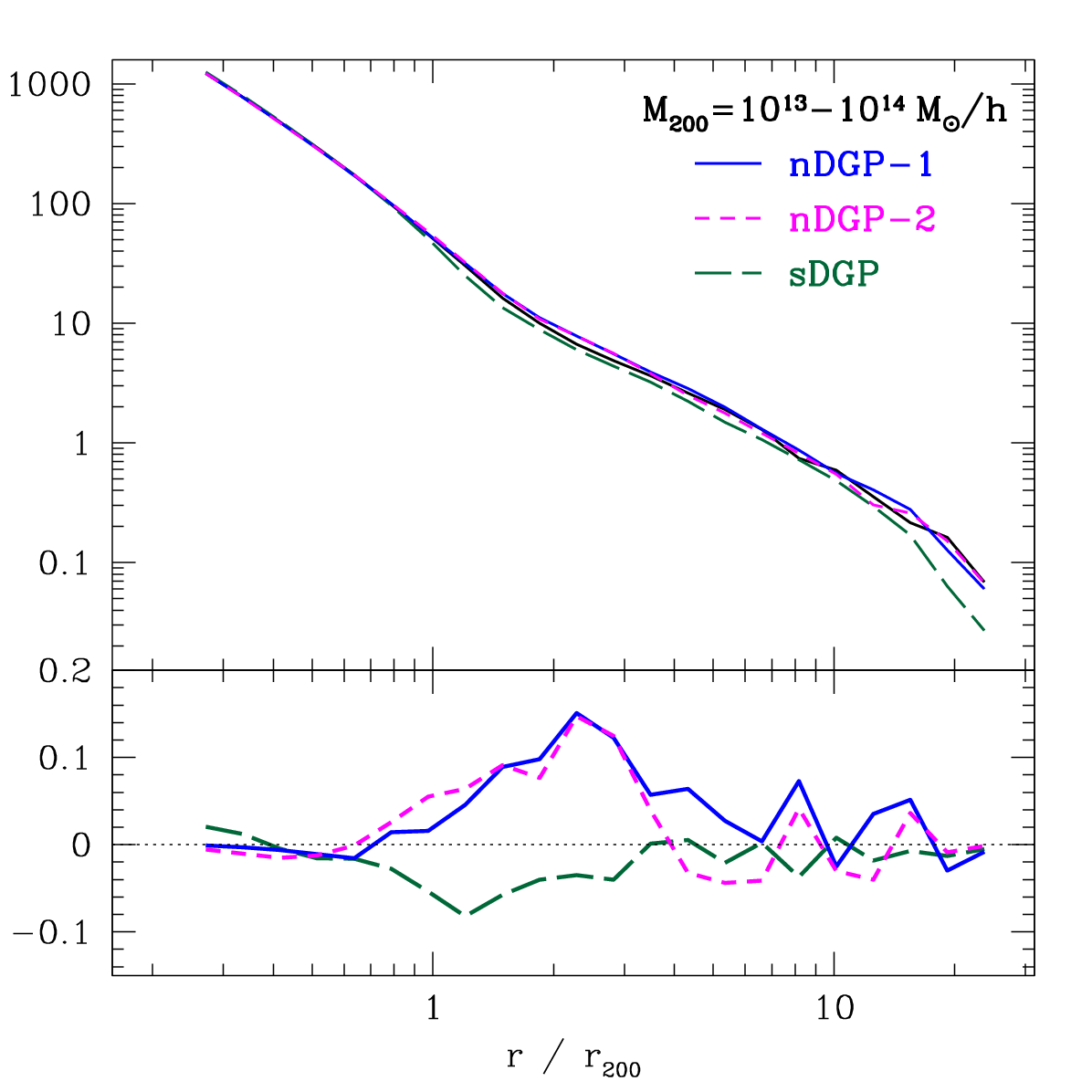}
\includegraphics[width=0.32\textwidth]{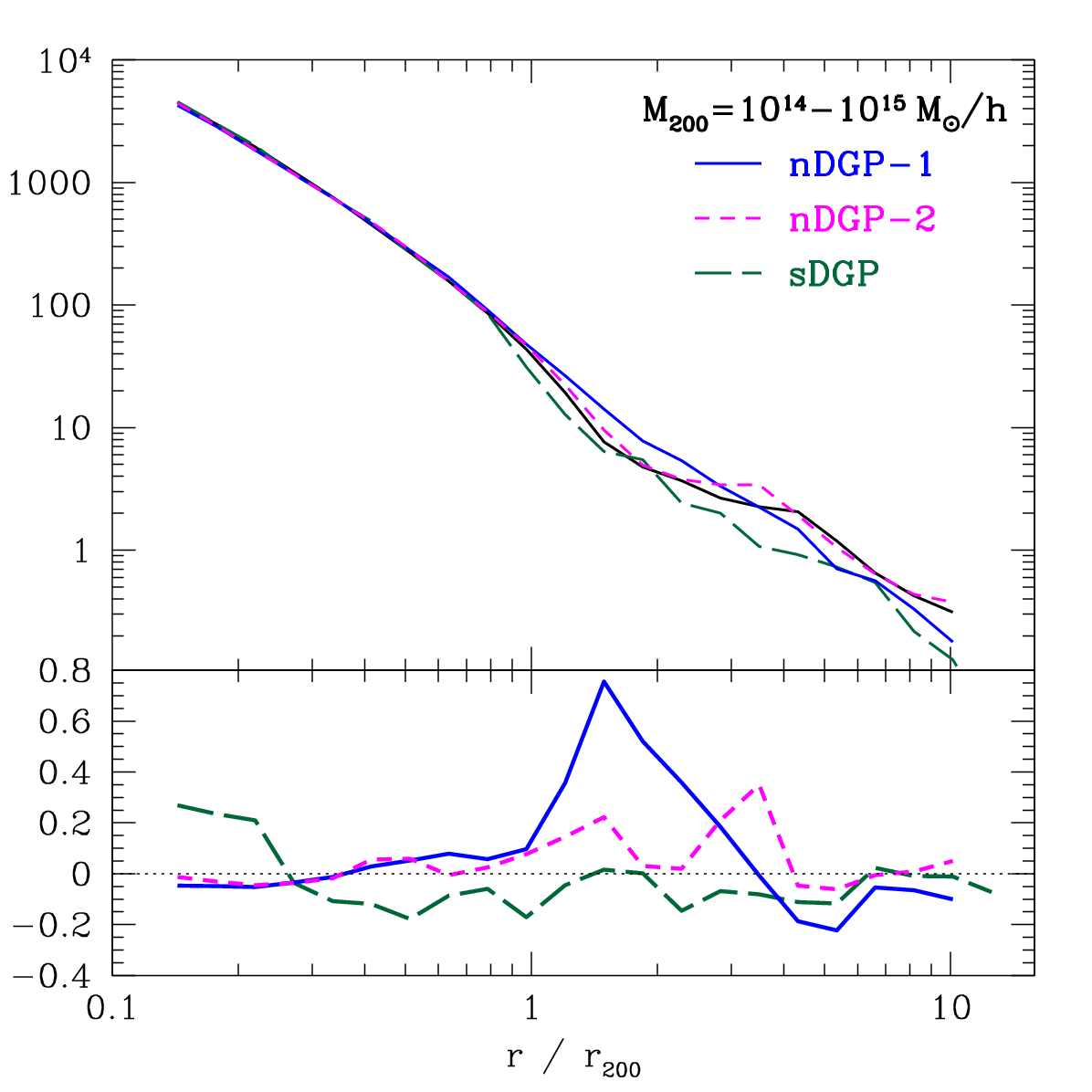}
\caption{{\small Average halo density profiles measured in the full DGP simulations.
Each figure corresponds to a fixed mass range: $\lg M/\Msunh = 12-13$
(left), $13-14$ (middle), $14-15$ (right). The top panels show profiles of
$\d_\rho \equiv (\rho-\rhob)/\rhob$ vs. $r$ in units of $r_{200}$, while
the bottom panel shows the relative deviation from the GR simulation with
the same expansion history in each case ($\L$CDM for nDGP, QCDM for sDGP).
For all figures, individual halo profiles were scaled to their respective
$r_{200}$ before averaging.}
\label{fig:rhoprof}}
\end{figure}

Once dark matter halos have been identified, we can study the distribution
of mass around them. We stack the radially averaged density profiles of halos in 
mass intervals as described in \cite{HPMhalopaper}. The center-of-mass of
each halo is calculated as the center of mass of particles within $1.4\:r_{\rm cell}$
of the center of the central (highest density) cell, which comprise
a large fraction of the halo mass.
In order to reduce scatter
within the mass bin, we scale each density profile to its own $r_{200}$ before 
stacking. We then bootstrap over all halos in the given
mass range in order to determine the average profile.
The spatial resolution of our particle-mesh simulations is limited by the
fixed size of grid cells $r_{\rm cell}$ (see Tab.~\ref{tab:runs}). 
We measure halo profiles down to the grid scale, though we
expect that profiles have converged only at scales of several grid cells.  
We only use the highest resolution boxes ($\Lbox=64\Mpch$) for 
the profile measurements.

\reffig{rhoprof} (upper panels) shows the halo density profiles measured in the full DGP
simulations, in three mass bins: $10^{12}-10^{13}\Msunh$ (left),
$10^{13}-10^{14}\Msunh$ (center), $10^{14}-10^{15}\Msunh$ (right).
In the inner regions, $r \lesssim r_{200}$, all DGP simulations show
the same, universal profile. Moreover, the profiles match those of the
corresponding GR+DE simulations, as shown in the lower panels in \reffig{rhoprof}.
This is because the inner regions of halos are assembled early on in the 
history of the universe, where the effects of the brane-bending mode are
very small ($\beta\gg 1$). Though our simulations cannot probe deep into the halos,
it seems unlikely there will be significant departures from GR at still smaller
radii.

Outside of the virial radius of halos, where matter is still infalling and tidal
fields are important, modified gravity effects can be seen: the departures
from the GR simulations typically peak around $2-3\:r_{200}$, and decrease again
towards larger radii. As expected,
they are positive for the nDGP+DE simulations, largest for nDGP--1, and
negative for sDGP. Note that the profiles for the highest mass bin are noisy
due to small halo statistics (especially in the case of sDGP where the abundance
of massive halos is significantly suppressed). For $r \gtrsim r_{200}$,
\reffig{rhoprof} should be interpreted as a halo-mass correlation function (scaled to
$r_{200}$), which is related to the mass correlation function and power spectrum.
Hence, the enhancement of the halo-mass correlation function in the nDGP simulations,
peaking at a few $r_{200}$, is another way of seeing the power spectrum
enhancement, peaking at $k\sim 0.7\iMpch$ (\reffig{Pk}).

\begin{figure}[t!]
\centering
\includegraphics[width=0.48\textwidth]{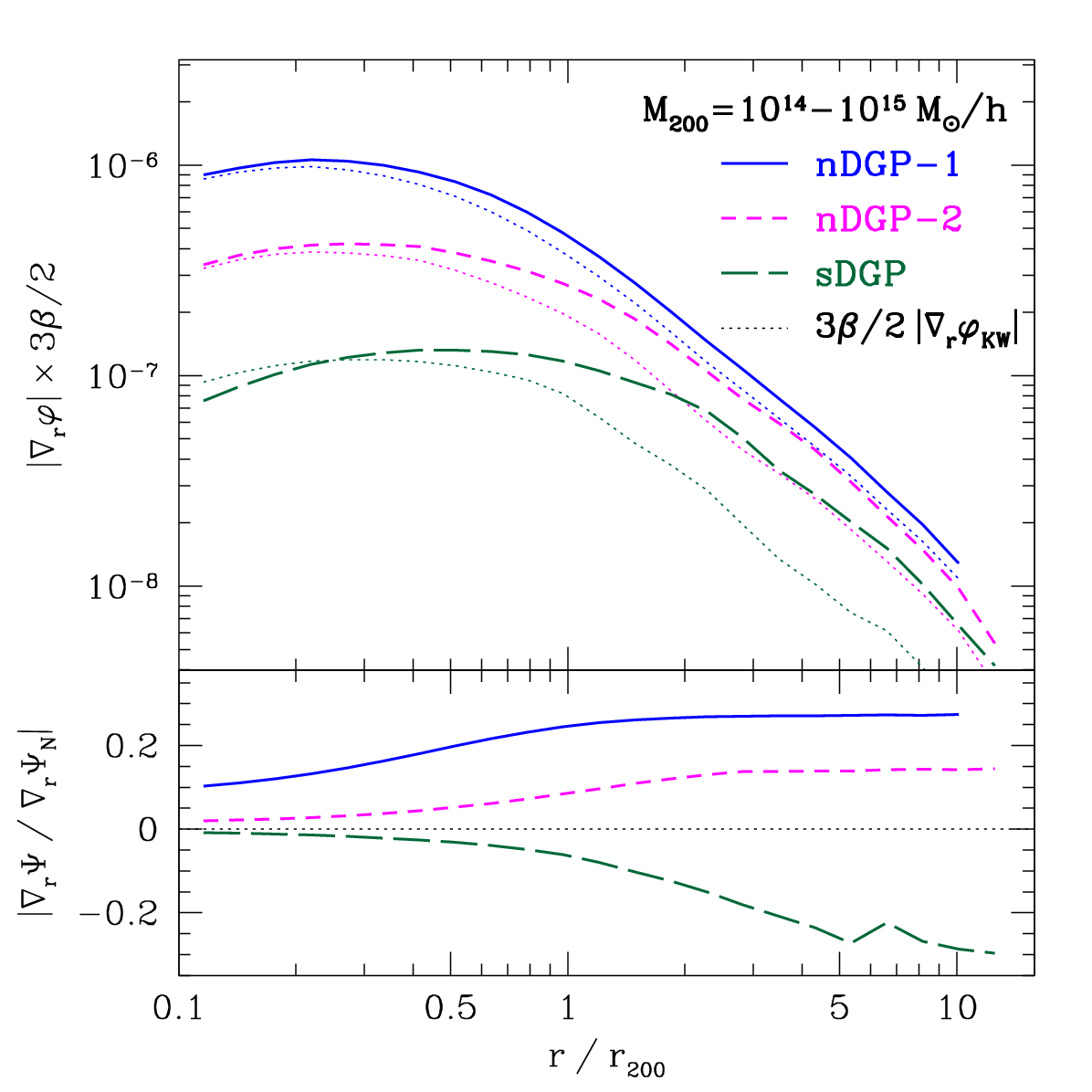}
\caption{{\small \textit{Top panel:} Average profiles of the radial gradient
of $\ph$, i.e. the force modification, measured in the full DGP simulations
and scaled by $3\beta/2$, for halos of
mass $\lg M/\Msunh = 14-15$. The profiles were averaged in a similar way
as the density profiles.
\textit{Bottom panel:} Deviation of the dynamical potential $\Psi$ [\refeq{Psi}]
from the Newtonian potential or lensing potential $\Phi_-=\Psi_N$ [\refeq{PsiN}]
 around the same dark matter halos. This quantity is probed
by Solar System tests.}
\label{fig:gradphi}}
\end{figure}

We can also look at the behavior of the brane-bending mode within halos.
\reffig{gradphi} (top panel) shows the average profile of the radial gradient
$|\nabla_r\ph|\times 3\beta/2$ measured in the full DGP simulations 
around the most massive (and best-resolved) halos. We show the gradient
of $\ph$ as the quantity with observable effects, since $\ph$ itself is
not observable. 
The prefactor is chosen so that the quantity approaches $\nabla_r\Phi_-$ on
linear scales (\refsec{linear}). Clearly, $\ph$ and its gradient are suppressed
within halos. As expected, the suppression is most severe for sDGP,
and weakest for nDGP--2. In each case, the thin dashed lines in \reffig{gradphi} show
the approximate solution for $\ph$ presented in \cite{KW}, calculated
given the density field as described in \cite{DGPMpaper} (see also \refapp{KW}). While the approximation
works well at the very center of the halos, it does not describe the solution
at $r \gtrsim 0.5\: r_{200}$. Moreover, the discrepancy becomes larger
the stronger the non-linearities in the $\ph$ equation are, i.e. the
larger $r_c$ is.

The lower panel of \reffig{gradphi} shows the gradient of the dynamical
potential around the same halos compared to the gradient of the Newtonian
or lensing potential $\Psi_N = \Phi_-$ which obeys the usual, unmodified
Poisson equation. A discrepancy between the dynamical potential of a given mass
and its lensing potential is precisely the quantity that is constrained by
Solar System tests, described via the post-Newtonian
parameter $\gamma$ \cite{Will}. \reffig{gradphi} shows that
this deviation is indeed suppressed within massive halos in our simulations,
although the simulations lack the resolution to follow the suppression very
far into the cores of halos. As expected, the suppression is strongest
for sDGP which has the largest non-linearity parameter $g$ (\refsec{vain}),
and weakest for nDGP--1.

\subsection{Bispectrum}
\label{sec:bis}

\begin{figure}[t!]
\centering
\includegraphics[width=0.48\textwidth]{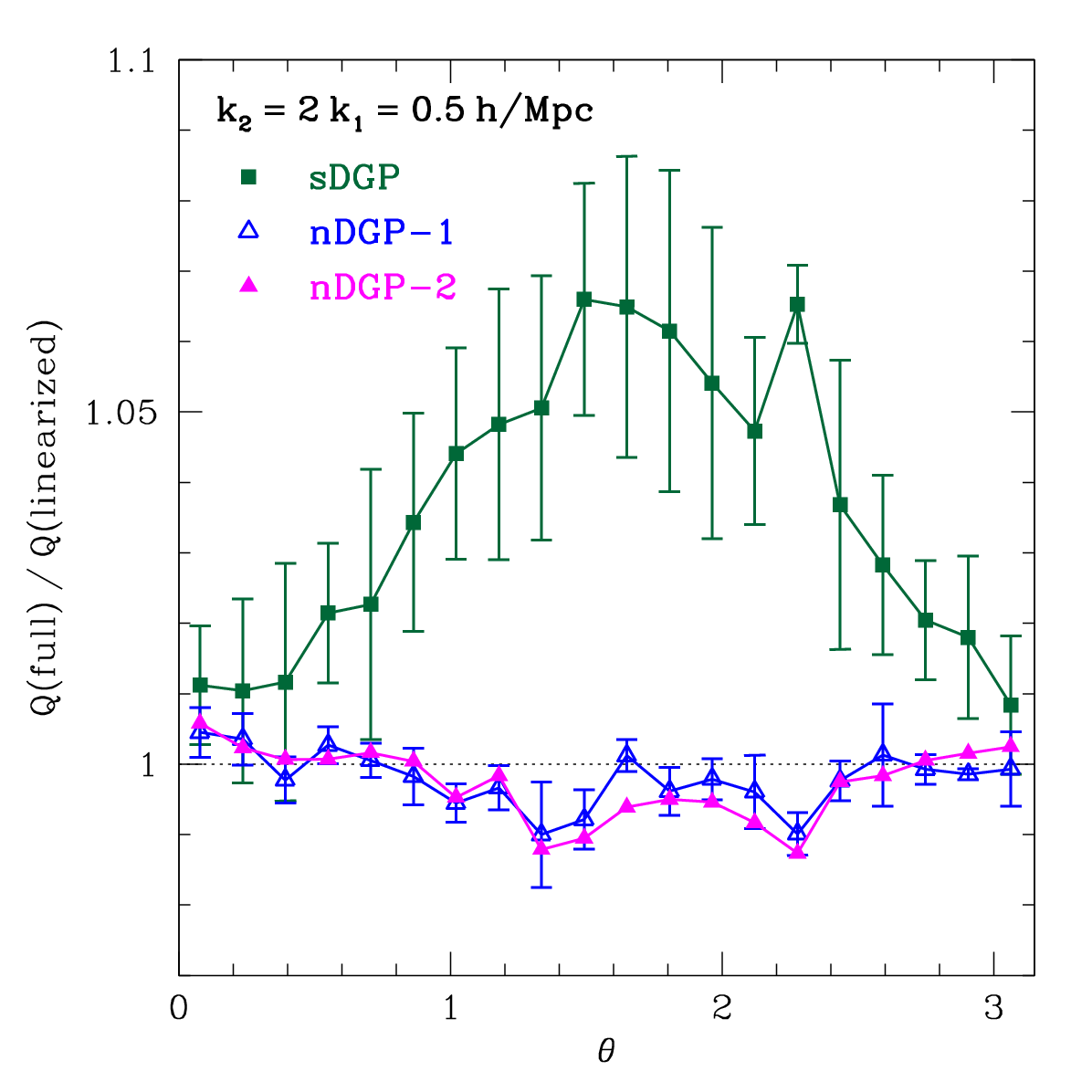}
\caption{{\small Ratio of the reduced bispectrum measured in full and
linearized DGP simulations, vs the angle between the wavevectors
$\v{k}_1=0.5\iMpch$, $\v{k}_2=1\iMpch$. 
This is a direct probe of the configuration dependence of the Vainshtein effect.
For each model, the bispectra were 
measured in the 256$\Mpch$ boxes. The errors are bootstrap error bars on 
the ratio determined from the different runs. We have omitted the error bars
on the nDGP--2 model for clarity - they are very similar to those for nDGP--1.}
\label{fig:bis}}
\end{figure}

The matter bispectrum in DGP is, in principle, a probe of the
quadratic non-linearity in the brane-bending mode equation. This
non-linearity adds an additional contribution to the tree-level bispectrum
generated by the ordinary gravitational non-linearities \cite{ScI}. As discussed
in \refsec{vain}, the specific quadratic self-interactions of the brane-bending
mode lead to a dependence on geometry. In Fourier space, the non-linearity
in \refeq{phiQS} leads to a coupling between two Fourier modes $\tilde\ph(\vk_1)$,
$\tilde\ph(\vk_2)$ with a kernel proportional to $1 - (\hat k_1\cdot\hat k_2)^2$,
where $\hat k_i$ denote unit vectors of the respective wavenumbers. As shown
in \cite{ScI}, this structure
manifests itself as a characteristic change of the shape dependence of the
bispectrum, which is commonly expressed in terms of the reduced bispectrum
$Q$:
\be
Q(\vk_1,\vk_2,\vk_3) \equiv \frac{B(\vk_1,\vk_2,\vk_3)}{P(\vk_1)P(\vk_2) + 
P(\vk_2)P(\vk_3) + P(\vk_3)P(\vk_1)},
\ee
where $B(\vk_1,\vk_2,\vk_3)$ is the bispectrum. In order to separate 
these characteristic effects of the brane-bending mode interactions as cleanly as
possible from the ordinary gravitational non-linearities, we compare the bispectrum
measured in the full DGP simulations to that of the linearized DGP simulations,
which have an identical linear growth factor, and hence are very similar in
their state of non-linear evolution. \reffig{bis} shows the
ratio of the reduced bispectrum measured in the two cases, for fixed wavenumbers
of $k_1=0.25\iMpch$ and $k_2=0.5\iMpch$, as a function of the angle
between them, $\cos\theta = \hat k_1\cdot \hat k_2$. The bispectrum
was measured in the simulations using Monte Carlo integration in a very similar
way as described in \cite{ScEtal98}.
We again average the ratios run-by-run, and estimate error bars by bootstrapping over
realizations.

The strongest effects are seen for the sDGP case, since this cosmology has the 
largest $r_c$ and hence the lowest threshold for the onset of the $\ph$ self-interactions.
Note that the bispectrum is enhanced in the full sDGP simulations for equilateral configurations
$\theta \approx \pi/2$, 
while it is very close to the linearized simulations for squeezed configurations
$\theta = 0,\;\pi$. 
For squeezed configurations, which correspond to planar geometry, the kernel 
$1 - (\hat k_1\cdot\hat k_2)^2 = 0$ and the non-linearities vanish. For equilateral configurations,
the suppression of $\ph$ is strongest, which leads to an enhanced
bispectrum in the full DGP simulations since the brane-bending mode
is repulsive. For nDGP--1
and nDGP--2, we see a hint of the opposite effect: here $\ph$ is attractive, and
a suppression due to the self-interactions in equilateral configurations leads to a suppression
of the bispectrum relative to the linearized simulations. The overall effect is much
smaller, since $r_c$ is smaller and the non-linearities in $\ph$ only become important
at higher densities (i.e., smaller scales).

The result for the sDGP bispectrum agrees very well with that found in \cite{ScII}.
Note that lacking very large box sizes, we measure the bispectrum at relatively small scales.
The ordinary
gravitational non-linearities which enter at higher order already contribute at the percent level
on the scales accessible in our simulations.
Hence, we do not attempt a quantitative comparison with tree-level perturbation theory
predictions here.

\section{Discussion}
\label{sec:concl}

We have introduced the nDGP+DE model based on the normal branch of DGP
with a smooth dark energy component on the brane, which results in
an expansion history that is precisely $\Lambda$CDM. 
This model can serve as an approximation, especially on sub-horizon scales,
to more general braneworld
models whose cosmological solutions have not been obtained yet.
In addition, the $r_c\rightarrow \infty$ limit precisely corresponds
to General Relativity plus cosmological constant.

Geometric measurements such as the acoustic scale constrained from the CMB,
BAO, and Supernovae observations do not constrain the crossover scale $r_c$
between 4D and 5D gravity in this scenario. The growth of large-scale structure
however does offer a sensitive probe of braneworld gravity via the effects
of the brane-bending mode which mediates an additional attractive force.
In the case of the self-accelerating branch, the brane-bending
mode is repulsive, so that the modified gravity effects are sign-flipped.

In order to study the formation of structure in these models, it is necessary
to solve the non-linear equations of the brane-bending mode in conjunction
with the ordinary gravitational dynamics in N-body simulations \cite{DGPMpaper}. These non-linear
interactions, which are responsible for the Vainshtein mechanism \cite{Vainshtein72,DeffayetVainshtein02} 
restoring General Relativity in high density environments, also leave characteristic
signatures in the bispectrum in the simulations \cite{ScI}, which we have
found in both sDGP and nDGP simulations. In addition, the non-linearities
manifest themselves in a suppression of the brane-bending mode around massive
halos.

The matter power spectrum in nDGP+DE shows a characteristic enhancement 
compared to that of $\Lambda$CDM, increasing towards smaller scales up to
$k\sim 0.7\iMpch$, and decreasing on even smaller scales. This is evidenced
in a complementary way by the profiles of dark matter halos and their environments
(halo-mass correlation function). Similar to what was found in $f(R)$ simulations
\cite{HPMhalopaper}, the inner regions of halos are not affected by the
enhanced forces since they assembled at early times. The strongest effects are
seen in the near environment of halos, around $2-3\:r_{200}$, corresponding to
the scale where the power spectrum enhancement peaks.

The abundance of massive halos is known to be a sensitive probe of
the growth of structure \cite{WhiteEtal93,EkeEtal98,BorganiEtal01,HPMhalopaper}. 
We found that indeed massive clusters are $2-4$ times more numerous for
the nDGP+DE model with $r_c=500-3000\:$Mpc. Current cluster abundance
measurements (e.g. \cite{Vikhlinin,RozoEtal09,MantzEtal09})
should be able to constrain the crossover
scale $r_c$ to be at least of order Gpc - a constraint which can be placed independently
of the specific DGP expansion history.

As the next step in understanding the effects of braneworld gravity on large scale
structure, an upcoming paper will detail a model of the matter power spectrum and halo
mass function based on spherical collapse and the halo model.
Such a model can also serve as an efficient way of dealing with
cosmological parameter dependencies when comparing with actual 
observations \cite{fRcluster}. In the future, by means of these simulations
and a physical model of their results, we will hopefully be able to probe 
the next generation of braneworld scenarios via their effect on the growth
of cosmic structure.

\acknowledgments
We would like to thank Wayne Hu, Kazuya Koyama, Roman Scoccimarro,
Mark Wyman, and Justin Khoury for
discussions and insightful comments. We thank the Aspen Center for Physics
where part of this work was completed for hospitality.

The simulations used in this work have been performed on the Joint 
Fermilab - KICP Supercomputing Cluster, supported by grants from Fermilab,
Kavli Institute for Cosmological Physics, and the University of Chicago. 
This work was supported by the Kavli Institute for Cosmological 
Physics at the University of Chicago through grants NSF PHY-0114422 and 
NSF PHY-0551142, and by the Gordon and Betty Moore Foundation at Caltech.

\appendix

\section{Comparison with $\bm{G_{\rm eff}(\d)}$ approximation}
\label{app:KW}

We now discuss the approximation adopted in the simulations of \cite{KW} and 
compare it with the results of our full solution of the brane-bending mode
equation~(\ref{eq:phiQS}). Consider a spherical ``tophat'' mass with a constant density
contrast $\d\rho$. Then, the ansatz $\ph(r) = A\:r^2 + C$ solves \refeq{phiQS},
and in particular we have:
\be
(\nabla_i\nabla_j\ph)^2 = \frac{1}{3}(\nabla^2\ph)^2.
\ee
For this special case, the non-linear terms combine to an equation of
$\ph$ which is algebraic in $\nabla^2\ph$:
\be
\nabla^2\ph + \frac{2 r_c^2}{9\beta\:a^2}(\nabla^2\ph)^2 = \frac{8\pi G a^2}{3\beta}\rhob\d.
\ee
This can then be solved for $\nabla^2\ph$ in terms of a non-analytic function
of the overdensity $\d$:
\bea
\nabla^2\ph &=& 8\pi G_{\rm eff}(\d)\,a^2\bar\rho_m \delta,\label{eq:phiKW}\\
G_{\rm eff} &=& \frac{2}{3\beta}\frac{\sqrt{1+\eps}-1}{\eps},\quad
\eps \equiv \frac{8 H_0^2r_c^2}{9\beta^2}\Om a^{-3} \delta.\label{eq:Geff}
\eea
Thus, the brane-bending mode is determined in this approximation by
a Poisson equation with density-dependent gravitational constant $G_{\rm eff}(\d)$.
Since this is a linear equation in $\ph$ which can be solved via Fourier transform, 
a simulation using this approximation is similar in terms of computing time 
as ordinary GR simulations, and much less computationally expensive than 
solving the full non-linear differential $\ph$ equation.

We now discuss the caveats of this ``$G_{\rm eff}$ approximation''.
For more general density profiles $\d\rho\neq$~const., \refeq{phiKW} is only
an approximate solution. More importantly, as already
pointed out in \cite{KW}, the approximation does not have the correct
large-distance behavior even for an isolated top-hat mass, so that scales 
in the linear regime of cosmological perturbations can in principle be 
affected by this incorrect behavior.

To see this, consider an isolated mass with $\d\rho = 0$ for $r$ larger
than some radius $R$. Integrating \refeq{phiQS} over a sphere with radius
$r > R$, we see that the two non-linear terms in \refeq{phiQS}
cancel via partial integration, leaving only boundary terms which become
increasingly suppressed as we let $r \rightarrow\infty$. Since the
right-hand side is just proportional to the enclosed mass $M$ within $r$,
we see that $\ph$ approaches the linearized solution 
$\ph(r) = (2/3\beta) G M/r$ at large $r$, irrespective of the strong
non-linearities close to the mass. For this reason, we recover the
linear predictions of the DGP model (\refsec{linear}) on large scales
in the simulations.

In contrast, integrating \refeq{phiKW} over the sphere with radius $r>R$,
we see that the asymptotic behavior in this approximation is
$\ph(r)\rightarrow (2/3\beta) \overline{G_{\rm eff}} M/r$, where 
$\overline{G_{\rm eff}}$
is the effective gravitational constant averaged over the mass. As 
\refeq{Geff} shows, $G_{\rm eff} \ll G$ if the density contrast $\d\gg 1$.
Hence, the far field of $\ph$ is strongly suppressed in this approximation.
In particular, if most of the cosmological mass is in non-linear structures
with $\d > 1$, we do not expect to recover linear theory predictions on
large scales in this approximation.
\begin{figure}[t!]
\centering
\includegraphics[width=0.48\textwidth]{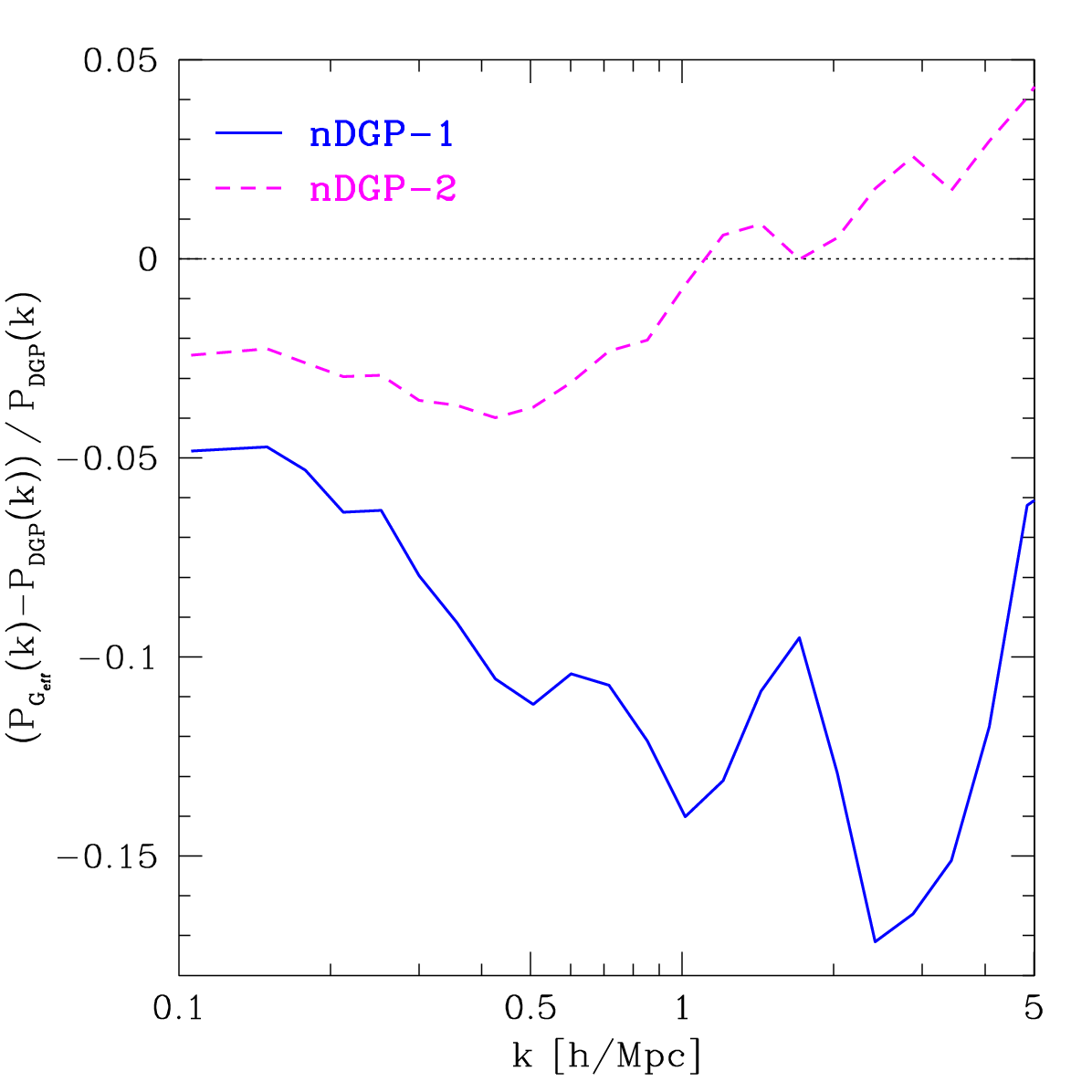}
\caption{{\small Relative deviation of the power spectrum at $z=0$ measured
in a simulation using the $G_{\rm eff}(\d)$ approximation, $P_{G_{\rm eff}}$,
from that of the full DGP simulation, $P_{\rm DGP}(k)$ for nDGP--1 and nDGP--2.
In each case, results are shown for a ``typical'' run with $\Lbox=64\Mpch$.}
\label{fig:KW}}
\end{figure}

As this discussion shows, the error encurred with the $G_{\rm eff}(\d)$ approximation in cosmological N-body
simulations depends on the resolution of the simulations: very low resolution
simulations (with large box sizes) which resolve little non-linear 
structure will not be affected; high-resolution simulations which
resolve very small dark matter halos will show a strong suppression of
the $\ph$ field and the corresponding modified forces on large scales, if
the $G_{\rm eff}$ approximation is used. In order to test how big
the effect is in our moderate resolution simulations, we reran a representative
simulation of our smallest box ($\Lbox=64\Mpch$) for the nDGP--1 and nDGP--2 
models using \refeq{phiKW} instead of \refeq{phiQS}.  For this, we chose a 
realization with a power spectrum close to the average of our 6 realizations
of the $64\Mpch$ box.  \reffig{KW} shows the relative
deviation in the power spectrum at $z=0$ in simulations with the 
$G_{\rm eff}$ approximation from that of the full DGP simulation with the same
initial conditions, for nDGP--1 and nDGP--2.  The deviations are around 
5--10\% for nDGP--1 and $\sim 3$\% for nDGP--2.  The magnitude of deviations
is noticeable given our $\lesssim 1$\% precision on the 
power spectrum deviation
on quasilinear scales (\reffig{Pk}).  As expected from our discussion, the
deviations persist on the largest scales probed by this box size.
Furthermore, power is suppressed in the $G_{\rm eff}$ simulations due to the
suppression of the attractive $\ph$-mediated force.  The deviations are
larger in nDGP--1 simply due to the stronger effect of $\ph$ on structure
formation in the small-$r_c$ model.
Thus, while the use of the $G_{\rm eff}(\d)$ approximation would not have a dramatic
impact on the main results presented in this paper, care must be taken in 
future high-resolution simulations of braneworld models: the artefacts of 
this approximation will increase rather than shrink with increasing resolution.

\clearpage
\bibliography{DGPM}

\end{document}